\documentclass[fleqn,usenatbib,useAMS]{mnras}

\usepackage{graphicx}	% Including figure files
\usepackage{amsmath}	% Advanced maths commands
\usepackage{amssymb}	% Extra maths symbols
\usepackage{multicol}        % Multi-column entries in tables
\usepackage{bm}		% Bold maths symbols, including upright Greek
\usepackage{pdflscape}	% Landscape pages
\usepackage{array}

\usepackage[T1]{fontenc}
\usepackage{ae,aecompl}

\usepackage{newtxtext,newtxmath}
\DeclareRobustCommand{\VAN}[3]{#2}
\let\VANthebibliography\thebibliography
\def\thebibliography{\DeclareRobustCommand{\VAN}[3]{##3}\VANthebibliography}

\newcommand{\Msun}{\rm M_{\odot}}

\newcommand{\HII}{{H}{\sc {ii}}}
\newcommand{\mm}{$\mu$m}

\title[Radio spectral properties of star-forming galaxies and their impact on the FIR-radio correlation]{Radio spectral properties of star-forming galaxies in the MIGHTEE-COSMOS field and their impact on the far-infrared-radio correlation}

\author[Fangxia An et al.]{
Fangxia An,$^{1}$\thanks{E-mail: fangxia@idia.ac.za, fangxiaan@gmail.com}
M. Vaccari,$^{1,2}$
Ian Smail,$^{3}$
M.J. Jarvis,$^{4,5}$
I. H. Whittam,$^{4,5}$
C. L. Hale,$^{6}$
S. Jin,$^{7,8}$
\newauthor
J. D. Collier,$^{9,10}$
E. Daddi,$^{11}$
J. Delhaize,$^{12}$
B. Frank,$^{9}$
E. J. Murphy,$^{13}$
M. Prescott,$^{1}$
S. Sekhar,$^{9,14,1}$
\newauthor
A. R. Taylor,$^{9,1}$
Y. Ao,$^{15,16}$
K. Knowles,$^{17,18}$
L. Marchetti,$^{12,2}$
S. M. Randriamampandry,$^{19,20}$
\newauthor
Z. Randriamanakoto$^{19}$
\\
$^{1}$Inter-University Institute for Data Intensive Astronomy, and Department of Physics and Astronomy, University of the Western Cape, Robert Sobukwe Road,\\ 7535 Bellville, Cape Town, South Africa\\
$^{2}$INAF-Istituto di Radioastronomia, via Gobetti 101, 40129 Bologna, Italy\\
$^{3}$Centre for Extragalactic Astronomy, Department of Physics, Durham University, Durham, DH1 3LE, UK\\
$^{4}$Astrophysics, University of Oxford, Denys Wilkinson Building, Keble Road, Oxford, OX1 3RH, UK\\
$^{5}$Department of Physics and Astronomy, University of the Western Cape, Robert Sobukwe Road, 7535 Bellville, Cape Town, South Africa\\
$^{6}$School of Physics and Astronomy, The University of Edinburgh, Institute for Astronomy, Royal Observatory, Blackford Hill, Edinburgh EH9 3HJ, UK\\
$^{7}$Instituto de Astrofísica de Canarias (IAC), E-38205 La Laguna, Tenerife, Spain\\
$^{8}$Universidad de La Laguna, Dpto. Astrofísica, E-38206 La Laguna, Tenerife, Spain\\
$^{9}$Inter-University Institute for Data Intensive Astronomy, and Department of Astronomy, University of Cape Town, Private Bag X3, Rondebosch 7701, South Africa\\
$^{10}$School of Science, Western Sydney University, Locked Bag 1797, Penrith, NSW 2751, Australia\\
$^{11}$CEA, IRFU, DAp, AIM, Universit{\'e} Paris-Saclay, Universit{\'e} Paris Diderot, Sorbonne Paris Cit{\'e}, CNRS, F-91191 Gif-sur-Yvette, France\\
$^{12}$Department of Astronomy, University of Cape Town, Private Bag X3, Rondebosch 7701, South Africa\\
$^{13}$National Radio Astronomy Observatory, 520 Edgemont Road, Charlottesville, VA 22903, USA\\
$^{14}$National Radio Astronomy Observatory, 1003 Lopezville Road, Socorro, NM 87801, USA\\
$^{15}$Purple Mountain Observatory and Key Laboratory for Radio Astronomy, Chinese Academy of Sciences, 8 Yuanhua Road, 210034, Nanjing, China\\
$^{16}$School of Astronomy and Space Science, University of Science and Technology of China, Hefei, Anhui, 230026, China\\
$^{17}$Department of Physics and Electronics, Rhodes University, PO Box 94, Makhanda 6140, South Africa\\
$^{18}$South African Radio Astronomy Observatory, 2 Fir Street, Observatory, Cape Town 7405, South Africa\\
$^{19}$South African Astronomical Observatory, PO Box 9, Observatory 7935, Cape Town, South Africa\\
$^{20}$A\&A, Department of Physics, Faculty of Sciences, University of Antananarivo, B.P. 906, Antananarivo 101, Madagascar\\
}
\date{Accepted XXX. Received YYY; in original form ZZZ}

\pubyear{2021}

\begin{document}
\label{firstpage}
\pagerange{\pageref{firstpage}--\pageref{lastpage}}
\maketitle

\begin{abstract}
We study the radio spectral properties of 2,094 star-forming galaxies (SFGs) by combining our early science data from the MeerKAT International GHz Tiered Extragalactic Exploration (MIGHTEE) survey with VLA, GMRT radio data, and rich ancillary data in the COSMOS field. These SFGs are selected at VLA 3\,GHz, and their flux densities from MeerKAT 1.3\,GHz and GMRT 325\,MHz imaging data are extracted using the ``super-deblending'' technique. The median radio spectral index is $\alpha_{\rm 1.3GHz}^{\rm 3GHz}=-0.80\pm0.01$ without significant variation across the rest-frame frequencies $\sim$1.3--10\,GHz, indicating radio spectra dominated by synchrotron radiation. On average, the radio spectrum at observer-frame 1.3--3\,GHz slightly steepens with increasing stellar mass with a linear fitted slope of $\beta=-0.08\pm0.01$, which could be explained by age-related synchrotron losses. Due to the sensitivity of GMRT 325\,MHz data, we apply a further flux density cut at 3\,GHz ($S_{\rm 3GHz}\ge50\,\mu$Jy) and obtain a sample of 166 SFGs with measured flux densities at 325\,MHz, 1.3\,GHz, and 3\,GHz. On average, the radio spectrum of SFGs flattens at low frequency with the median spectral indices of $\alpha^{\rm 1.3GHz}_{\rm 325MHz}=-0.59^{+0.02}_{-0.03}$ and $\alpha^{\rm 3.0GHz}_{\rm 1.3GHz}=-0.74^{+0.01}_{-0.02}$. At low frequency, our stacking analyses show that the radio spectrum also slightly steepens with increasing stellar mass.  
By comparing the far-infrared-radio correlations of SFGs based on different radio spectral indices, we find that adopting $\alpha_{\rm 1.3GHz}^{\rm 3GHz}$ for $k$-corrections will significantly underestimate the infrared-to-radio luminosity ratio ($q_{\rm IR}$) for $>$17\% of the SFGs with measured flux density at the three radio frequencies in our sample, because their radio spectra are significantly flatter at low frequency (0.33--1.3\,GHz).
\end{abstract}

\begin{keywords}
radio continuum: galaxies -- methods: observational -- galaxies: formation --Galaxy: evolution
\end{keywords}

%%%%%%%%%%%%%%%%%%%%%%%%%%%%%%%%%%%%%%%%%%%%%%%%%%

%%%%%%%%%%%%%%%%% BODY OF PAPER %%%%%%%%%%%%%%%%%%

\section{Introduction} \label{sec:intro}
The radio emission from star-forming galaxies (SFGs) consists of non-thermal synchrotron radiation from the cosmic-ray (CR) electrons spiralling in the interstellar magnetic field, and thermal free-free emission produced by the electrostatic interactions between charged particles, mostly free electrons and ions, in \HII\ regions. In SFGs, the CR electrons are primarily accelerated by shocks associated with supernovae remnants. In addition, the radio continuum emission at $\nu \la 30$\,GHz is unaffected by dust attenuation. These characteristics make radio emission a key observable that provides information about the CR electrons, magnetic fields, photoionization rate, dust-obscured star-formation, etc., in galaxies \citep[see][for a review]{Condon92}. Therefore, deep radio continuum surveys are essential for our understanding of physical processes at work in SFGs from the local to the distant Universe \citep[e.g.,][]{Ibar09,Murphy09, Beswick15,Prandoni15, Jarvis16, Smolcic17a, Ocran20a}.

In the last four decades, numerous studies have focused on characterising the radio continuum spectra of SFGs to explore the relative contributions of synchrotron and free-free emission, components or structures corresponding to these two radiation mechanisms, and their correlation with emission in other wavebands  \citep[e.g.,][]{Klein88, Richards98, Haarsma00,Condon02,Seymour08, Williams10, Marvil15, Tabatabaei17, Klein18,Gim19}. Traditionally, the radio spectrum of a SFG is assumed to be a superposition of a steep synchrotron spectrum with spectral index $\alpha=-0.8$ ($S_{\nu} \propto \nu^{\alpha}$) and a flat free-free spectrum with $\alpha=-0.1$, which is also supported by observations of some nearby SFGs with a star-formation rate (SFR) $<10\,\Msun$\,year$^{-1}$ \citep[e.g.,][]{Condon92,Niklas97, Tabatabaei17}. However, studies of luminous and ultraluminous infrared galaxies (ULIRGs), starburst galaxies, dwarf galaxies, or even some normal SFGs in the local Universe and at high redshift have shown that the radio continuum spectral energy distributions (SEDs) of SFGs are rarely characterised well by a single power-law \citep[e.g.,][]{Clemens10, Williams10, Murphy13, Marvil15, Klein18,Tisanic19,Thomson19}.

A well-determined radio spectrum for SFG is critically important for studies that are based on the rest-frame radio power, especially those at high redshift, which are most sensitive to the assumed $k$-corrections. These include studies of the correlation observed between radio and far-infrared (FIR) emission of SFGs, which is widely adopted to calibrate the SFR measured from radio continuum emission, and the definition of the ``radio excess'' that is used to classify active galactic nuclei (AGN) and SFGs \citep[e.g.,][]{Helou85, Yun01, Bell03, Ivison10, Mao11,Del13,Magnelli15,Hindson18,Ocran20a, Ocran20b, Ocran21}. The FIR-radio correlation (FIRRC) has been well established for local SFGs at GHz frequencies. However, for SFGs at high redshift, the radio emission at rest-frame $\sim$1.4\,GHz is shifted to low frequencies ($\nu_{\rm obs}<1$\,GHz at $z>0.4$). Therefore, the appropriate spectral index that is used to $k$-correct the observed radio flux densities at $\sim$1.4\,GHz for SFGs at $z>0$ should be drawn from lower frequency ($\nu_{\rm obs}<1.4$\,GHz) data. For SFGs, previous work demonstrated that at $\nu<1$\,GHz, the \HII\ region becomes optically thick and the relativistic bremsstrahlung, synchrotron self-absorption, and Razin effects become more important than that at high frequency \citep[e.g.,][]{Rybicki79, Condon92, Lacki13, Chyzy18}. These effects can suppress the radio emission and hence flatten the radio spectrum at low frequency, as has been observed in nearby SFGs \citep[e.g.,][]{Condon92,Clemens10,Murphy13, Marvil15,Kapinska17,Chyzy18}. 

For the high-redshift Universe, although some recent studies also found a flatter radio spectrum at low frequency, these works only focus on the high-SFR \citep[SFR\,$>$\,100\,$\Msun$\,year$^{-1}$,][]{Tisanic19} or submillimeter bright galaxies \citep{Thomson19}. In addition, due to the lack of high-quality radio data at low frequency, most of the studies based on rest-frame radio spectra of SFGs at high redshift have adopted a spectral index measured from high frequency ($\nu_{\rm obs}=$\,1--10\,GHz), or even assumed a fixed spectral slope, when $k$-correcting the observed flux densities \citep[e.g.,][]{Appleton04,Ivison10,Ibar10,Sargent10, Mao11, Delhaize17,Algera20, Delvecchio21}. However, some recent works suggest that these assumptions will dramatically overestimate the radio flux density at low frequency and thus strongly affect the FIRRC of SFGs \citep[e.g.,][]{Schleicher13, Delhaize17, Galvin18, Gim19}. Therefore, high-quality radio observations at both low and high frequency from the Square Kilometre Array (SKA) \footnote{https://www.skatelescope.org/} and its precursors, such as MeerKAT and ASKAP, will be extremely important in accurately determining the radio spectra of SFGs at high redshift and then shedding light on galaxy formation and evolution through the radio window \citep{Beswick15,Prandoni15, Jarvis16}. 

In this work, we use the early science data from the MeerKAT International GHz Tiered Extragalactic Exploration (MIGHTEE) survey and take advantage of existing radio data from {\it Karl G. Jansky} Very large Array (VLA) and Giant Metrewave Radio Telescope (GMRT) surveys, as well as the rich ancillary data in the COSMOS field to select SFGs up to $z\sim3$ and study their radio spectral properties. By studying the shape of radio spectra with the physical properties derived from other bands, we discuss the mechanisms that characterise the radio spectra at low and high frequencies and their impact on the FIRRC of SFGs. 

The radio and other ancillary data used in this work are described in Section $\S$\ref{s:observation}. We present our analyses of radio spectral properties based on these data in Section $\S$\ref{s:Analysis}. Our results are shown in Section $\S$\ref{s:results}. We discuss and summarise our results in Section $\S$\ref{s:discussion} and $\S$\ref{s:conclusion} respectively. Throughout this paper, we adopt the AB magnitude system \citep{Oke74} and assume a flat $\Lambda$CDM cosmological model with the Hubble constant $H_0 = 67.27$\,km\,s$^{-1}$\,Mpc$^{-1}$, matter density parameter $\Omega_{\rm m} = 0.32$, and cosmological constant $\Omega_{\Lambda} = 0.68$ \citep{Planck16}.

\section{Observations and data} \label{s:observation}

\subsection{MIGHTEE 1.3\,GHz survey in the COSMOS field} \label{s:obs_mightee}
The MIGHTEE survey is one of the MeerKAT large survey projects\footnote{http://public.ska.ac.za/meerkat/meerkat-large-survey-projects}. The main purpose of MIGHTEE is to obtain deep GHz radio continuum, spectral line, and polarisation observations to study the cosmic evolution of galaxies and AGN. Details of the science goals and observation plan of the MIGHTEE survey are described in \cite{Jarvis16}.  Taking advantage of existing multi-wavelength data, the MIGHTEE project is surveying four well-studied extragalactic fields, i.e., COSMOS (1\,deg$^2$), ELAIS S1 (1.6\,deg$^2$), CDFS (8.3\,deg$^2$) and XMM-LSS (6.7\,deg$^2$) fields, with a median sensitivity of 2\,$\mu$Jy\,beam$^{-1}$ at L-band (900--1670\,MHz) and 1\,$\mu$Jy\,beam$^{-1}$ at S-band (1750--3500\,MHz) with the latter only for the CDFS and the COSMOS fields. We refer the reader to \cite{Jarvis16} for more details. 

In this work, we use the early science data of the COSMOS field (R.A.:\,10:00:28.6, Decl.:\,$+$02:12:21), which were observed in 2018 and 2020 with the MeerKAT L-band receiver ($\nu_{\rm c}=1.3$\,GHz). Because of the frequency dependence of the primary beam as well as the wide bandwidth, the effective frequency gradually decreases from the centre outwards. However, the MIGHTEE-COSMOS early science data were observed with a single pointing and the decrease of the effective frequency is $<0.04$\,GHz within the primary beam ($\sim$1\,deg FWHM). This variation is negligible in studying radio spectra using MeerKAT, VLA 3\,GHz, and GMRT 325\,MHz data. Therefore, we use $\nu_{\rm c}=1.3$\,GHz as the effective frequency of MeerKAT L-band data in this work. 

In total, the observations include 17.45 hours on source for the central pointing in the COSMOS field. Details of the observations and data reduction will be presented in I.\,Heywood et al.\ in preparation but see also a summary in \cite{Delhaize21}. Here we provide a brief overview of the final reduced early science data. 

Because high-sensitivity and high-resolution cannot be achieved at the same time, the MIGHTEE continuum data are imaged twice with Briggs' robust parameter 0.0 and $-$1.2. In this work, to obtain a less-biased sample of SFGs for the study of their radio spectral properties (Section $\S$\ref{s:results}), we only use the high-sensitivity imaging data. The maximum-sensitivity image reaches a thermal noise of 1.7\,$\mu$Jy\,beam$^{-1}$ with an angular resolution of 8$\farcs6\times8\farcs6$. In addition to the thermal noise, the contribution from the numerous faint and unresolved astronomical sources, namely classical confusion, increases the mean root-mean-square (RMS) to 4.5\,$\mu$Jy\,beam$^{-1}$.

\subsection{Other radio data in the COSMOS field} \label{sec:obs_ts}
\subsubsection{VLA 3\,GHz and 1.4\,GHz data}
The {\it Karl G. Jansky} VLA-Cosmic Evolution Survey (VLA-COSMOS) 3\,GHz Large Project mapped the entire 2\,deg$^{2}$ COSMOS field with a median RMS of 2.3\,$\mu$Jy\,beam$^{-1}$ and an angular resolution of 0$\farcs$75 \citep{Smolcic17a}. In total, 10,830 radio sources were detected at $>5\,\sigma$. We refer the reader to \cite{Smolcic17a} for more details. 

The VLA 1.4\,GHz data are from the VLA-COSMOS Large Project \citep{Schinnerer07} and the VLA-COSMOS Deep Project \citep{Schinnerer10}. The former covered the entire 2\,deg$^{2}$ COSMOS field with a mean RMS noise of 15\,$\mu$Jy\,beam$^{-1}$. The latter combined the existing data from the VLA-COSMOS Large Project and an additional VLA L-band observations of the central 50$\arcmin\times$50$\arcmin$ subregion in the COSMOS field to reach 1\,$\sigma$ sensitivity $\sim$12\,$\mu$Jy\,beam$^{-1}$ and angular resolution $1\farcs5\times1\farcs4$. In total, 2,856 radio sources with signal-to-noise ratio (SNR)\,$>5$ (peak flux densities) were detected from this combined imaging data. Details are presented in \cite{Schinnerer10}.

\subsubsection{GMRT 610\,MHz and 325\,MHz data}
The GMRT 325\,MHz data used in this work are from \cite{Tisanic19}. Observations were carried out with GMRT under the project 07SCB01 (PI: C.\,Crof),  which covered the entire $\sim$2\,deg$^{2}$ COSMOS field and reached a median RMS of 97\,$\mu$Jy\,beam$^{-1}$ and an angular resolution of 10$\farcs$8$\times$9$\farcs$5. In total, 633 radio sources were detected at $>5\,\sigma$. Details are presented in \cite{Tisanic19}.

\cite{Tisanic19} also published GMRT 610\,MHz data, which has a median RMS of 39\,$\mu$Jy\,beam$^{-1}$. However, as described in \cite{Tisanic19}, we find that for sources having measured flux densities at GMRT 325\,MHz and 610\,MHz, MeerKAT 1.3\,GHz, and VLA 1.4\,GHz and 3\,GHz, there is a deficit in flux density computed at GMRT 610\,MHz when compared with flux densities at the other four frequencies. However, it is not a systematic underestimation. Therefore, a simple 20\% correction for all sources as employed by some previous work \citep[e.g.,][]{Kolokythas15} is not appropriate. \cite{Tisanic19} suggests that this flux density offset might be positionally dependent, but they found that simple exclusion of pointings did not yield a specific pointing which produces these offsets. Since this offset will dramatically affect the measurement of radio spectral index, we exclude the GMRT 610\,MHz data in the following analyses. 

\subsection{Optical/NIR/FIR/submillimeter catalogues} \label{sec:catalogue}
The COSMOS field is one of the largest extragalactic fields with deep multi-wavelength data. Previous work have described in detail the available dataset in the COSMOS field from X-ray to ultraviolet (UV), optical, near-infrared (NIR), mid-infrared (MIR), FIR, (sub)millimeter and radio bands \citep[e.g.,][]{Scoville07,Capak07,Sanders07, Ilbert09, Brusa10, Civano12, Civano16, Laigle16, Liu19, Simpson19, Simpson20}. In this work, except the radio data, we do not use these multi-wavelength data directly but utilise the published multi-wavelength catalogues in the COSMOS field. Here we give a short introduction of these catalogues.

The optical/NIR photometric catalogue used in this work is the COSMOS2015 catalogue, which is primarily based on the UltraVISTA-DR2 surveys \citep{Laigle16}. The COSMOS2015 catalogue includes 1,182,108 sources detected from a $\chi^{2}$ sum of $YJHK_{\rm s}$ and $z^{++}$ imaging data. To match the photometry from Near-UV (NUV) to MIR, \cite{Laigle16} computed the total fluxes estimated from the corrected 3$\arcsec$ aperture flux, at 31 photometric bands from NUV to MIR, including 12 medium bands. Using the photometrically matched data, \cite{Laigle16} fitted the SEDs from NUV to MIR for these sources and estimated their photometric redshift, stellar mass, SFR, rest-frame luminosities, and so on. For more details, see \cite{Laigle16}. 

The FIR catalogue used in this work is a ``super-deblended'' FIR to (sub)millimeter photometric catalogue from \cite{Jin18}. They utilised the positions of  UltraVISTA $K_{\rm s}$- and/or VLA 3\,GHz-detected sources \citep{Laigle16, Smolcic17a} to obtain the point spread function (PSF) fitted flux densities from MIPS 24\,$\mu$m images \citep{Le09} and VLA 1.4\,GHz and 3\,GHz images \citep{Schinnerer10,Smolcic17a}. They then select sources with SNR\,$>3$ at radio or 24\,$\mu$m as priors. With an additional ``mass-selected'' sample, \cite{Jin18} adopt a ``super-deblending'' technique developed by \cite{Liu18} to ``deblend'' the FIR to (sub)millimeter photometry from {\it Spitzer} \citep{Le09}, {\it Herschel} \citep{Oliver10, Lutz11, Bethermin12}, SCUBA2 \citep{Cowie17, Geach17}, AzTEC \citep{Aretxaga11}, and MAMBO \citep{Bertoldi07} images in the COSMOS field for a total 221,428 sources. Using the ``super-deblended'' photometry, \cite{Jin18} fitted FIR to millimeter SEDs for these sources and estimated their SFR by integrating 8--1000\,\mm\ infrared luminosities ($L_{\rm IR}$) derived from the best-fit SEDs. In this work, we also use the stellar mass from the ``super-deblended" catalogue in our analyses (Section $\S$\ref{s:results}), which are from \cite{Laigle16} and \cite{Muzzin13}, and was estimated by using the best-fit optical to NIR SED with a Chabrier IMF \citep{Chabrier03}.

The two catalogues of ALMA-detected submillimeter galaxies (SMGs) in the COSMOS field from \cite{Liu19} and \cite{Simpson20} are not used in the main analyses of this work but are used in marking this extreme population in our sample of SFGs to compare with previous studies \citep{Tisanic19, Thomson19}. The A$^{3}$COSMOS\footnote{https://sites.google.com/view/a3cosmos} catalogue includes 1,134 ALMA sources with SNR\,$\ge5.4$ extracted from all publicly available ALMA archive data in the COSMOS field \citep{Liu19}. The AS2COSMOS catalogue published in \cite{Simpson20} includes 260 ALMA SMGs detected in observations of an essentially complete sample of 184 SCUBA-2 sources brighter than $S_{\rm 850}=6.2$\,mJy \citep{Simpson19}.

\section{Analysis}\label{s:Analysis}
To investigate the spectral properties of radio-detected sources, we first measure the radio flux densities at different frequencies from VLA, MeerKAT, and GMRT surveys in the COSMOS field. We also describe the selection of SFGs in this section.

\subsection{The ``super-deblended'' flux densities }\label{s:super-deblended_a}

As described in Section $\S$\ref{sec:catalogue}, \cite{Jin18} adopted a ``super-deblending'' technique developed by \cite{Liu18} to ``deblend''  FIR to (sub)millimeter photometries of sources selected from NIR or radio. The purpose of the ``super-deblending'' technique is to deal with source confusion caused by the low angular resolution of FIR and single-dish (sub)millimeter imaging data when cross-matching them with optical/NIR data. Since the ``super-deblended''  catalogue we use in this work is from \cite{Jin18}, we take it as an example to shortly describe the basic steps of ``super-deblending'' technique. Firstly, \cite{Jin18} used the position of high-resolution $K_{\rm s}$-band- or VLA 3\,GHz-detected sources to perform PSF fitting in MIPS 24\,$\mu$m images from \cite{Le09} and VLA 1.4 and 3\,GHz images from \cite{Schinnerer10} and \cite{Smolcic17a} respectively by using the G{\scriptsize{ALFIT}} \citep{Peng02, Peng10}. For high SNR sources (SNR\,$>$10 at 24\,$\mu$m and SNR\,$>$20 at VLA 1.4\,GHz and 3\,GHz), they run a second pass fitting allowing for up to one and two pixel variations of prior source positions for 24\,$\mu$m and radio images respectively. This returns a cleaner residual image compared to the results of first-pass fitting. Secondly, they performed a Monte Carlo simulation in MIPS 24\,$\mu$m, 1.4\,GHz, and 3\,GHz maps respectively to verify potential flux biases and calibrate the uncertainties of photometric measurements. \cite{Liu18} and \cite{Jin18} compared the ``super-deblended'' flux densities with directly measured ones from previous works \citep[e.g.,][]{Le09,Schinnerer10, Smolcic17a} and confirmed the reliability of ``super-deblending'' technique. We refer readers to \cite{Liu18} and \cite{Jin18} for more details.

In this work, we aim to match the photometry of the MeerKAT data at 1.3\,GHz with those of VLA 1.4\,GHz and 3\,GHz, and GMRT 325\,MHz data. Because of the low resolution of GMRT and MeerKAT data as well as $\sim10\times$ difference in the angular resolutions between them and VLA 3\,GHz data, we also adopt the ``super-deblending'' technique to measure flux densities at these radio frequencies. To ``deblend'' the MeerKAT image, we adopt the same $K_{\rm s}$+\,3\,GHz prior sample in \cite{Jin18} and obtain their ``super-deblended'' flux densities at 1.3\,GHz. However, because of the low sensitivity of GMRT data, we only use 3\,GHz detection as priors in ``deblending'' 325\,MHz images. This will not affect the sample size of this study as it is based on a radio-selected sample.

%
% Figure 1
%
\begin{figure}
\centering
\includegraphics[width=0.48\textwidth]{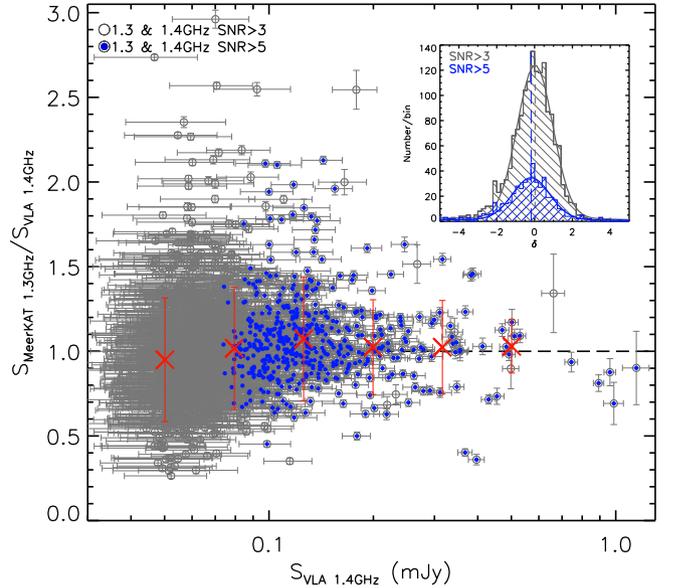}
\caption{Ratio of ``super-deblended'' flux densities at MeerKAT 1.3\,GHz and VLA 1.4\,GHz as a function of 1.4\,GHz flux density for the 1,599 sources with SNR\,$>$\,3 at both frequencies. We mark the sources that have SNR\,$>$\,5 at both frequencies by blue solid points. 
The red crosses are the median $(S_{\rm 1.3} / S_{\rm 1.4})$ for sources within each 1.4\,GHz flux density bin.
The inset plot shows the distribution of normalized difference of ``super-deblended'' flux densities at these two frequencies for sources with SNR\,$>$\,3 and SNR\,$>$\,5 at both frequencies respectively. Overall, despite the scatter, the ``super-deblended'' flux densities extracted from the VLA 1.4\,GHz and MeerKAT 1.3\,GHz data are consistent with each other within the uncertainties for sources with SNR\,$>$\,5 if we consider their slightly different effective frequencies.}
\label{f:check_vla_mightee_flux}
\end{figure}

To test the self-consistency of the ``super-deblending'' technique, we compare the ``super-deblended'' flux densities at 1.3\,GHz ($S_{1.3}$) from MIGHTEE survey and at 1.4\,GHz ($S_{1.4}$) from VLA-COSMOS survey for the 1,599 sources with SNR\,$>$\,3 at both frequencies as shown in Figure~\ref{f:check_vla_mightee_flux}. We limit our analyses within the area of the MeerKAT primary beam (1\,deg diameter) to avoid low sensitivity regions at the edges of MIGHTEE survey. The relatively lower median value of $(S_{\rm 1.3} / S_{\rm 1.4})$ within the lowest 1.4\,GHz flux-density bin (Figure~\ref{f:check_vla_mightee_flux}) is caused by the very different sensitivities of the two datasets (Section~$\S$\ref{s:observation}). For the cut of SNR\,$>$\,3, the low sensitivity of VLA 1.4\,GHz data results in a relatively larger uncertainty of measured flux density at 1.4\,GHz (Figure~\ref{f:check_vla_mightee_flux}), which, in return, leads to a relatively higher flux density cut at VLA 1.4\,GHz. In contrast, the higher SNR cut (SNR\,$>$\,5) is less affected by this effect, as shown in Figure~\ref{f:check_vla_mightee_flux}. We therefore mainly rely on the sample with SNR\,$>$\,5 at both frequencies in verifying the self-consistency of the ``super-deblending'' technique.

The inset plot in Figure~\ref{f:check_vla_mightee_flux} shows the distributions of normalized difference between the two flux densities for the sources with SNR\,$>$\,3 and $>$\,5 respectively. The normalized difference is defined as: 
\begin{eqnarray} \label{e:equation0}
\delta = (S_{\rm 1.4}-S_{\rm 1.3})/\sqrt{(\sigma_{\rm 1.4}^{2} + \sigma_{\rm 1.3}^{2})} ,
\end{eqnarray}
where $\sigma$ refers to the uncertainty of ``super-deblended'' flux density. The mean differences are $\delta=0.02$ and $-0.18$ with the scatters of $\sigma_{\delta} = 0.94$ and 0.96 for the two subsamples respectively. The negative $\delta$ and the median $(S_{\rm 1.3} / S_{\rm 1.4}) = 1.04^{+0.02}_{-0.01}$ for sources with SNR\,$>$\,5 might be indicative of the slightly different effective frequencies of the two datasets, which results in $(S_{\rm 1.3} / S_{\rm 1.4}) = 1.06$ if we assume the spectral index $\alpha = -0.8$ between 1.3 and 1.4\,GHz. Therefor, although some of the sources appear to be outliers, the flux densities at these two frequencies are consistent with each other for the sources with SNR\,$>$\,5.

\subsection{SFGs selection}\label{s:super-deblended_SFGs}
Due to the high resolution and the high sensitivity of VLA 3\,GHz data, we select sources with SNR\,$>$\,5 at 3\,GHz from the ``super-deblended'' catalogue \citep{Jin18} to study the radio spectral properties in this work. The 3\,GHz is also the highest observer-frame frequency we use in studying radio spectrum in this work. Compared with selecting sources at low frequency, which bias the sample to the sources with a relatively steeper radio spectrum, the choice of 3\,GHz as the selection frequency provides a less-biased sample for studying the radio spectrum of SFGs in this work. In addition, the high sensitivity of MeerKAT 1.3\,GHz data guarantees a high completeness of studying radio spectrum at 1.3--3\,GHz for 3\,GHz-selected SFGs. We will come back to this in Section $\S$\ref{s:results}.

Within the area of the MeerKAT primary beam (1\,deg diameter), there are 4,509 VLA 3\,GHz sources with SNR\,$>$\,5. We first select the star formation dominated galaxies by removing  previously classified AGN or red galaxies from these radio sources. 

For X-ray AGN in the COSMOS field, \cite{Marchesi16} cross-correlated the {\it Chandra} X-ray detected sources from the {\it Chandra} COSMOS legacy Survey \citep{Civano16} with optical/NIR sources in the literature and identified optical/NIR counterparts for 97\% of 4,016 {\it Chandra} X-ray sources. We first cross-match VLA 3\,GHz radio sources with optical/NIR counterparts of X-ray sources from \cite{Marchesi16} by using a matching radius of 1$\arcsec$. We find that 566/4,509 radio sources have {\it Chandra} X-ray detections and 519 of them are classified as AGN in \cite{Marchesi16} according to their rest-frame X-ray luminosity ($L_{\rm X}>10^{42}$\,erg\,s$^{-1}$). We also cross correlate these 3\,GHz radio sources with optical/NIR counterparts of 2,012 X-ray sources from the {\it XMM-Newton} X-ray survey \citep{Cappelluti09,Brusa10} and find an additional six X-ray detected AGN.

The X-ray AGN-selection tends to miss low-luminosity AGN at high redshift, as they typically do not have accretion-related X-ray emission \citep{Marchesi16}. Therefore we also adopt the MIR color-color selection criteria in \cite{Donley12} for $z\le3$ sources. Since the ``super-deblended'' catalogue includes all VLA 3\,GHz  sources with SNR\,$>$\,5 \citep{Jin18}, we use the matched photometry from this catalogue and find that 2,760 sources in our sample have SNR\,$>$\,3 in all four IRAC bands. Among these, 240 meet the selection criteria of MIR AGN \citep{Donley12}. 

In the ``super-deblended" catalogue, \cite{Jin18} classified 726/4,509 radio-excess sources as radio-loud AGN. The criteria for radio excess are defined as ($S_{\rm OBS,radio}-S_{\rm SED,radio})/\sqrt{(\sigma^2_{\rm OBS,radio}+\sigma^2_{\rm SED,radio})}>3$ and $S_{\rm OBS,radio}>2\times S_{\rm SED,radio}$ \citep{Liu18}, where $S_{\rm OBS,radio}$ is the observed radio flux density and $S_{\rm SED,radio}$ is the flux density predicted by the best-fit FIR to millimeter SEDs (Section $\S$\ref{sec:catalogue}). For the 566 and 311 VLA 3\,GHz radio sources that are cross-matched with the optical/NIR counterparts of X-ray sources from \cite{Marchesi16} and from \cite{Brusa10}, 160 of them are identified as AGN according to their best-fit optical to NIR SED in \cite{Marchesi16}, and 392 and 183 are spectroscopically identified AGN in \cite{Marchesi16} and \cite{Brusa10} respectively. In addition, 3,958/4,509 radio sources are also included in the 5-$\sigma$ 3\,GHz catalogue presented in \cite{Smolcic17b}. Among them, 501, 271, 578, and 793 are classified as X-ray, MIR, SED, and radio excess AGN in \cite{Smolcic17b}. By integrating these classifications from the literature, in total we remove 1,916 AGN from the 4,509 VLA 3\,GHz radio sources. 

To further ensure a clean sample of SFGs, we also remove red galaxies with both $(M_{\rm NUV}-M_{r})>3.5$ and non-detection in the {\it Herschel} bands (100\,\mm, 160\,\mm, 250\,\mm, 350\,\mm, and 500\,\mm) from our sample, since previous work proved that this population shows typical properties that are consistent with those of radio AGN host galaxies \citep[e.g.,][]{Best06, Sadler14, Smolcic17b}. The absolute rest-frame {\it GALEX} NUV and Subaru $r$-band magnitudes are taken from the COSMOS2015 catalogue \citep{Laigle16}, while the non-detection by {\it Herschel} is defined as sources without $>$\,3\,$\sigma$ detection at any {\it Herschel} band. There are 761 galaxies which meet these selection criteria and 573 of them overlap with our AGN sample. Therefore, after removing 2,104 AGN and/or red galaxies, we obtain a sample of 2,405 SFGs with SNR\,$>$\,5 at VLA 3\,GHz. Among these, 2,094 have a photometric redshift or spectroscopic redshift in the ``super-deblended'' catalogue, in which the photometric redshift is from the COSMOS2015 catalogue \citep{Laigle16} while the spectroscopic redshift is from a new COSMOS master spectroscopic catalogue \citep[M. Salvato et al. in preparation;][]{Jin18}. 

%
% Figure 2
%
\begin{figure*}
\centering
\includegraphics[width=0.98\textwidth]{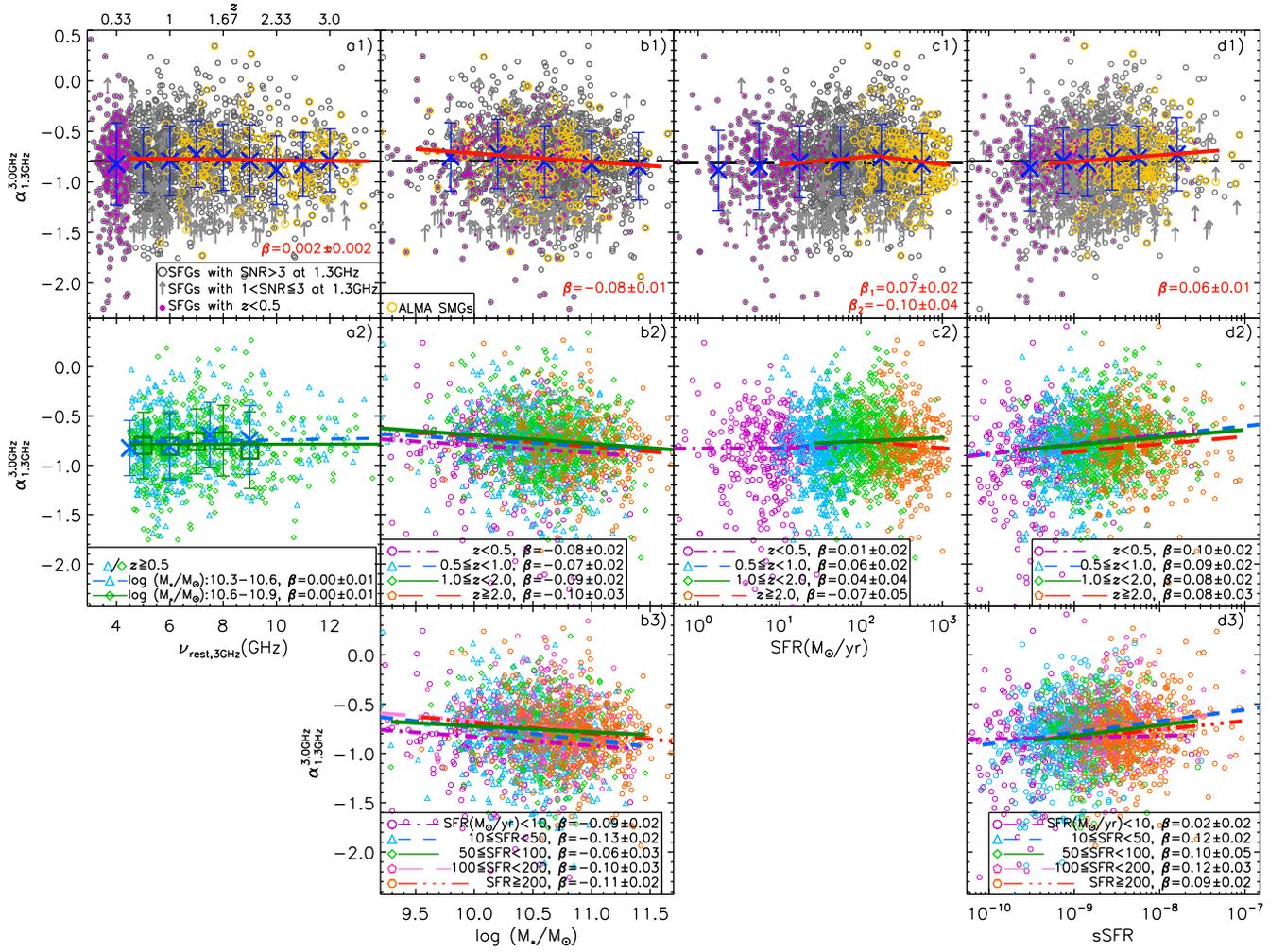}
\caption{Spectral index between the observer-frame frequencies of 1.3--3\,GHz, $\alpha^{\rm 3.0GHz}_{\rm 1.3GHz}$, as a function of rest-frame frequency corresponding to observed 3\,GHz, redshift (a1), stellar mass (b1), SFR (c1), and sSFR (d1) for the 2,012 VLA 3\,GHz-selected SFGs. The upward arrows show the lower limit of $\alpha^{\rm 3.0GHz}_{\rm 1.3GHz}$ for the 161 SFGs with 1\,$<$\,SNR\,$\le$\,3 at MeerKAT 1.3\,GHz. The black dashed line represents the median radio spectral index of $\alpha^{\rm 3.0GHz}_{\rm 1.3GHz}=-0.80\pm0.01$ for the remaining 1,851 SFGs with SNR\,$>3$ at MeerKAT 1.3\,GHz.
The blue cross or green square represents the median $\alpha^{\rm 3.0GHz}_{\rm 1.3GHz}$ of SFGs in each bin while the error bar shows the standard deviation. The slightly steeper spectrum at the lowest redshift (rest-frame), sSFR, and the two lowest SFR bins might be  biased by the missing flux of high-resolution VLA 3\,GHz observations for extended sources at low redshift. We therefore exclude galaxies with $z<0.5$ in the linear fits of the correlation between $\alpha^{\rm 3.0GHz}_{\rm 1.3GHz}$ and a physical property of the galaxies (red solid lines) to avoid the effect from missing flux of VLA 3\,GHz observations, although not all of low-redshift SFGs are affected. The linear fit is performed by minimizing the $\chi^{2}$ and using the uncertainties of $\alpha^{\rm 3.0GHz}_{\rm 1.3GHz}$ as the inverse weights. The slopes of these linear fits, $\beta$, are given on the right-bottom corner of each panel. We also mark the SFGs with ALMA detection (yellow circles) and find that they show the statistic properties consistent with the entire sample. For the SFGs with SNR\,$>3$ at MeerKAT 1.3\,GHz, we show the correlation between $\alpha^{\rm 3.0GHz}_{\rm 1.3GHz}$ and a physical property of the galaxies within stellar mass (a2), redshift (b2, c2, and d2) or SFR (b3, and d3) bins. The plot c2 suggests that the trend between SFR and $\alpha^{\rm 3.0GHz}_{\rm 1.3GHz}$ is caused by the redshift-dependent selection effect of our radio flux-limited sample. Therefore, the trend between radio spectral index and sSFR (d1) might be an another manifestation of the correlation between the radio spectrum and the stellar mass of SFGs. The plots in the first and second columns suggest that the trends between $\alpha^{\rm 3.0GHz}_{\rm 1.3GHz}$ and redshift or stellar mass are very unlikely to be caused by selection effect. Overall, on average, the radio spectral slope at observer-frame 1.3--3\,GHz is not strongly correlated with redshift/rest-frame frequency, but slightly steepens with increasing stellar mass and flattens with sSFR for radio-selected SFGs, although the scatter is large. 
}
\label{f:rest_3_z}
\end{figure*}

\section{Results} \label{s:results}
To study the radio spectral properties of SFGs, we first extract their flux densities from MeerKAT 1.3\,GHz, and GMRT 325\,MHz maps at their 3\,GHz positions by applying the ``super-deblending'' technique \citep{Jin18}. 

\subsection{Spectral index of SFGs between the observer-frame frequencies of 1.3--3\,GHz}\label{s:results_super}

For the 2,094 VLA 3\,GHz-selected SFGs, 96\% (2,012/2,094) of them have measured ``super-deblended'' flux densities with SNR\,$>$\,1 at MeerKAT 1.3\,GHz. Among them, 1,851 (92\%) have SNR\,$>$\,3 while the rest 161 have 1\,$<$\,SNR$\le$\,3. However, because of the low sensitivity of VLA 1.4\,GHz data, although 85\% (1,783/2,094) of these 3\,GHz-selected SFGs have the measured flux densities at VLA 1.4\,GHz, only a further 42\% (756/1,783) have SNR\,$>$\,3. Therefore, in this work, we use the ``super-deblended" flux densities at MeerKAT 1.3\,GHz and VLA 3\,GHz to measure the radio spectral index between the observer-frame frequencies of 1.3--3\,GHz for our selected SFGs. The radio spectral index, $\alpha_{\rm1.3GHz}^{\rm 3GHz}$, is measured by performing a linear fit in logarithmic space, and its uncertainty is inherited from the uncertainties of the flux densities. The flux densities at VLA 1.4\,GHz are included in the fit if the SFGs also have SNR\,$>$\,3 at VLA 1.4\,GHz.

There are 712 VLA 3\,GHz-selected SFGs with SNR\,$>$\,3 at both MeerKAT 1.3\,GHz and VLA 1.4\,GHz. We measure their radio spectral index between the observer-frame frequencies of 1.3--3\,GHz and obtain a median radio spectral index of $\alpha_{\rm1.3GHz}^{\rm 3GHz}=-0.80\pm0.01$ and $\alpha_{\rm1.3GHz}^{\rm 3GHz}=-0.79\pm0.01$ if we include or exclude the VLA 1.4\,GHz data in the measurement respectively. Here the error of the median spectral index is estimated from bootstrap resampling. The scatters of both measurements are $\sigma=0.31$. These results further suggest the self-consistency of ``super-deblending" technique (Section $\S$\ref{s:super-deblended_a}).

For the 2,012 SFGs that have measured flux densities at MeerKAT 1.3\,GHz (SNR\,$>$\,1), their median spectral index is $\alpha^{\rm 3.0GHz}_{\rm 1.3GHz}=-0.78\pm0.01$ with a scatter of $\sigma=0.37$. However, if we limit our sample to the SFGs with SNR\,$>$\,3 at MeerKAT 1.3\,GHz, we obtain a median spectral index of $\alpha^{\rm 3.0GHz}_{\rm 1.3GHz}=-0.80\pm0.01$ and a scatter of $\sigma=0.35$. Therefore, a higher SNR cut at MeerKAT 1.3\,GHz will bias our sample towards SFGs with a relatively steeper radio spectrum at observer-frame 1.3--3\,GHz, but we find that this effect is negligible because of the high sensitivity of MeerKAT 1.3\,GHz data. We thus study the correlations between $\alpha^{\rm 3.0GHz}_{\rm 1.3GHz}$ and physical properties of SFGs based on the sample of SFGs with SNR\,$>$\,3 at MeerKAT 1.3\,GHz as shown in Figure~\ref{f:rest_3_z}. For the 161 SFGs with 1\,$<$\,SNR\,$\le$\,3 at MeerKAT 1.3\,GHz, because of the relatively larger uncertainties of their 1.3\,GHz flux densities, we use the 5\,$\sigma$ limit of MeerKAT data to estimate the lower limit of their radio spectral indices (Figure~\ref{f:rest_3_z}).

In Figure~\ref{f:rest_3_z}, we show the measured spectral index between the observer-frame frequencies of 1.3--3\,GHz as a function of corresponding rest-frame frequency of VLA 3\,GHz, redshift, stellar mass, SFR, and specific star formation rate (sSFR) for the 2,012 SFGs. The sSFR is defined as the ratio of SFR and stellar mass. The relatively steeper spectrum at the lowest redshift, sSFR, and the two lowest SFR bins is likely to be caused by missing flux in the high-resolution VLA 3\,GHz observations. This results from the fact that for extended galaxies at low redshift, only part of their flux densities have been recovered in the high-resolution 3\,GHz observations. Therefore, we simply exclude the SFGs with $z<0.5$ in the linear fits shown in Figure~\ref{f:rest_3_z}.

As shown in the first panel (a1) of Figure~\ref{f:rest_3_z}, on average, the spectral index at observer-frame 1.3--3\,GHz is not strongly correlated with rest-frame frequency (therefore also redshift) of SFGs. The slope of the linear fit between $\alpha_{\rm 1.3GHz}^{\rm 3.0GHz}$ and rest-frame frequency ($\nu_{\rm rest, 3GHz}$) is $\beta_{(\alpha,\nu_{\rm rest,3GHz})}=0.002\pm0.002$.
We also show the trends between $\alpha^{\rm 3.0GHz}_{\rm 1.3GHz}$ and rest-frame frequency for SFGs with $10.3\le$\,log\,$(M_{*}/\Msun)<10.6$ and $10.6\le$\,log\,$(M_{*}/\Msun)<10.9$ respectively (plot a2 of Figure~\ref{f:rest_3_z}) and confirm that this trend is independent from the stellar mass of SFGs. Galaxies with $z<0.5$ are excluded in the plot a2 of Figure~\ref{f:rest_3_z} to avoid the effect from missing flux of VLA 3\,GHz observations. 

The second panel (b1) of Figure~\ref{f:rest_3_z} shows that the radio spectral slope at observer-frame 1.3--3\,GHz steepens slightly with increasing stellar mass with a linear fitted slope of $\beta=-0.08\pm0.01$. We find that this trend can be observed within different redshift or SFR bins as shown in the plot b2 and b3 of Figure~\ref{f:rest_3_z}. 

The third panel (c1) of Figure~\ref{f:rest_3_z} shows that, on average, the radio spectral slope first slightly flattens with increasing SFR but reverses from SFR\,$\ga$200\,$\Msun$\,yr$^{-1}$. The slopes of the linear fits are $\beta_{(\alpha,{\rm SFR\la200\,\Msun\,yr^{-1}})}=0.07\pm0.02$ and $\beta_{(\alpha,{\rm SFR\ga200\,\Msun\,yr^{-1}})}=-0.10\pm0.04$ respectively. However, the plot c2 of Figure~\ref{f:rest_3_z} suggests that this is caused by redshift-dependent selection effect of our radio flux-limited sample. These analyses suggest that the trend between $\alpha^{\rm 3.0GHz}_{\rm 1.3GHz}$ and sSFR ( $\beta_{(\alpha,{\rm sSFR})}=0.06\pm0.01$, plot  d1 of Figure~\ref{f:rest_3_z}) is to a large extent another manifestation of the correlation between $\alpha^{\rm 3.0GHz}_{\rm 1.3GHz}$ and stellar mass (b1 of Figure~\ref{f:rest_3_z}).

We also include SFGs with $z<0.5$ in the linear fits of the correlations between $\alpha^{\rm 3.0GHz}_{\rm 1.3GHz}$ and the physical properties shown in Figure~\ref{f:rest_3_z}. The slopes of these linear fits change to $\beta_{(\alpha,\nu_{\rm rest,3GHz})}=0.005\pm0.002$, $\beta_{(\alpha,{\rm log (M_{*}/\Msun)})}=-0.06\pm0.01$, $\beta_{(\alpha,{\rm SFR\la200\,\Msun\,yr^{-1}})}=0.05\pm0.01$, and $\beta_{(\alpha,{\rm sSFR})}=0.07\pm0.01$ respectively. These results suggest that the missing flux of high-resolution VLA 3\,GHz observations for extended sources at low redshift slightly affects the study of the correlation between $\alpha^{\rm 3.0GHz}_{\rm 1.3GHz}$ and a physical property of the SFGs in this work.

 We mark the 187 SFGs with ALMA detection from AS2COSMOS \citep{Simpson20} or A3COSMOS surveys \citep{Liu19} in Figure~\ref{f:rest_3_z} and find that their radio spectra at observer-frame 1.3--3\,GHz show the statistical properties consistent with the entire sample. 

Overall, out of the five physical properties shown in Figure~\ref{f:rest_3_z}, only the stellar mass and sSFR show robust correlations with the radio spectral index at observer-frame 1.3--3\,GHz of SFGs, although the scatter is large. We will discuss the physical mechanisms responsible for the trends between radio spectral index and these physical properties in Section $\S$\ref{s:discussion_high}.

\subsection{Spectral indices of SFGs between the observer-frame frequencies of 0.33--3\,GHz}\label{s:sed_super}
To study the radio spectral properties at low frequency, we include the GMRT 325\,MHz data to measure the radio spectral index between the  observer-frame frequencies of 0.33--3\,GHz for selected SFGs in the COSMOS field.

%
% Figure 3
%
\begin{figure}
\centering
\includegraphics[width=0.48\textwidth]{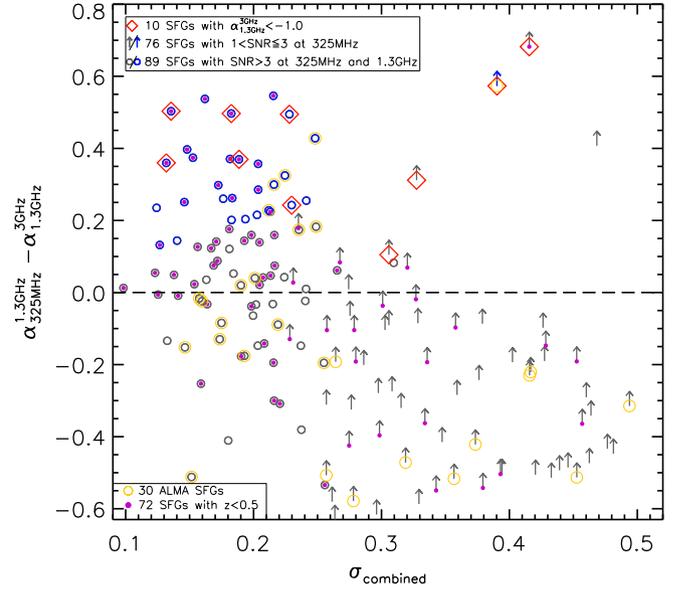}
\caption{Difference between radio spectral indices at low and high frequencies ($\alpha^{\rm 1.3GHz}_{\rm 325MHz}-\alpha^{\rm 3.0GHz}_{\rm 1.3GHz}$) as a function of their combined uncertainties for the 166 SFGs with $S_{\rm 3GHz}\ge50\,\mu$Jy and measured flux densities at the three radio frequencies. 
The upward arrows show the lower limit of ($\alpha^{\rm 1.3GHz}_{\rm 325MHz}-\alpha^{\rm 3.0GHz}_{\rm 1.3GHz}$) for the 76 SFGs with $1<$\,SNR\,$\le3$ at GMRT 325\,MHz.
The blue symbols are the 30 SFGs, which have a low-frequency radio spectrum significantly flatter than that at high frequency. We also mark the 72 SFGs with $z<0.5$ (purple points), the 30 ALMA SMGs (yellow circles), and ten SFGs with $\alpha^{\rm 3.0GHz}_{\rm 1.3GHz}<-1.0$ (red rhombs) respectively.}
\label{f:figure3}
\end{figure}

%
% Figure 4
%
\begin{figure*}
\centering
\includegraphics[width=0.96\textwidth]{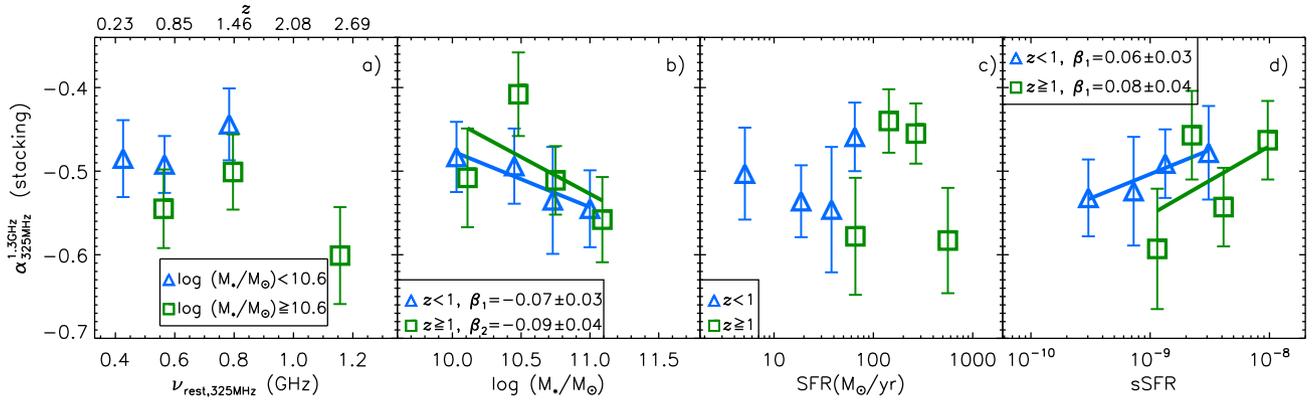}
\caption{Averaged $\alpha^{\rm 1.3GHz}_{\rm 325MHz}$, which is estimated by stacking MeerKAT and GMRT images for all of the 2,094 VLA 3\,GHz-selected  SFGs, as functions of rest-frame frequency corresponding to observed GMRT 325\,MHz, redshift, stellar mass, SFR, and sSFR. We divide the sample into the massive (log$(M_{*}/\Msun)\ge10.6$) and less-massive (log$(M_{*}/\Msun) <10.6$) subsamples when stacking SFGs within different redshift bins and find that, on average, the massive SFGs have a relatively steeper radio spectrum. We also divide our sample into high-redshift ($z\ge1$) and low-redshift ($z<1$) subsamples when stacking SFGs within different stellar mass, SFR, and sSFR bins. The results show that the averaged radio spectral slope at low frequency also steepens with increasing stellar mass, and flattens with increasing sSFR despite some fluctuations at high redshift ($z\ge1$).}
\label{f:radio_sindex_snr5_mightee_3GHz}
\end{figure*}

\subsubsection{Comparison of radio spectrum at high and low frequencies}\label{s:individual measure}

For the 2,094 VLA 3\,GHz-selected SFGs, we have measured the spectral index at observer-frame 1.3--3\,GHz for 96\% (2,012/2,094) of them (Section $\S$\ref{s:results_super}). However, because of the low sensitivity of the GMRT 325\,MHz data, only half of these 3\,GHz-selected SFGs have a measured ``super-deblended" flux density with SNR\,$>1$ at GMRT 325\,MHz. Therefore, we apply an additional flux density cut of $S_{\rm 3GHz}\ge50\,\mu$Jy to obtain a less-biased sample of SFGs for studying the radio spectral properties at observer-frame 0.33--3\,GHz. The criterion $S_{\rm 3GHz}\ge50\,\mu$Jy is chosen in order to balance the completeness of the SFGs with measured flux density at GMRT 325\,MHz and the sample size. Within the region for our analyses, there are 220 SFGs with $S_{\rm 3GHz}\ge50\,\mu$Jy. Among them, 91\% (201/220) and 84\% (184/220) have SNR\,$>1$ at MeerKAT 1.3\,GHz and GMRT 325\,MHz respectively. For the 19 sources that are brighter than 50\,$\mu$Jy at VLA 3\,GHz but lack a measured ``super-deblended" flux density at MeerKAT 1.3\,GHz, we visually inspect the MeerKAT 1.3\,GHz image and the residual map. We find that their residual maps are not clean after two rounds of G{\scriptsize{ALFIT}} PSF fitting \citep{Jin18}, which indicates that their flux density profiles cannot be well-approximated by the Gaussian distribution. For this reason, their MeerKAT 1.3\,GHz flux densities cannot be accurately derived by the ``super-deblending" technique. Likewise, of the 36 sources that lack the measured flux density with SNR\,$>1$ at GMRT 325\,MHz, four of them are also caused by the inability to fit with a Gaussian profile. In addition, the other 32 simply have SNR\,$<1$ at GMRT 325\,MHz.

We first measure the radio spectral indices between the observer-frame frequencies of 0.33--1.3\,GHz, 0.33--3\,GHz, and 1.3--3\,GHz for a sample of 166, 184, and 201 SFGs respectively, using the two flux densities at the endpoints of each frequency range. The median radio spectral indices are $\alpha^{\rm 1.3GHz}_{\rm 325MHz}=-0.59^{+0.02}_{-0.03}$, $\alpha^{\rm 3GHz}_{\rm 325MHz}=-0.65^{+0.02}_{-0.03}$, and 
$\alpha^{\rm 3.0GHz}_{\rm 1.3GHz}=-0.73^{+0.01}_{-0.02}$ with a scatter of $\sigma=0.29$, 0.19, and 0.22 respectively. Therefore, on average, the radio spectrum of SFGs flattens at low frequency. 

For the SFGs with $S_{\rm 3GHz}\ge50\,\mu$Jy and measured flux densities at MeerKAT 1.3\,GHz, 99\% (198/201) of them have SNR\,$>3$ at MeerKAT 1.3\,GHz. However, for the 184 SFGs with measured flux density at GMRT 325\,MHz, only 58\% (107/184) of them have SNR\,$>3$ at GMRT 325\,MHz. The measured radio spectral indices for a sample of 89, 107, and 198 SFGs with SNR\,$>3$ at the two corresponding frequencies are $\alpha^{\rm 1.3GHz}_{\rm 325MHz}=-0.69^{+0.03}_{-0.06}$, $\alpha^{\rm 3GHz}_{\rm 325MHz}=-0.74^{+0.01}_{-0.03}$, and $\alpha^{\rm 3.0GHz}_{\rm 1.3GHz}=-0.73^{+0.01}_{-0.02}$  with a scatter of $\sigma=0.19$, 0.13, and 0.22 respectively. Therefore, a higher SNR cut at GMRT 325\,MHz will bias our sample towards the SFGs with a relatively steeper radio spectrum at observer-frame 0.33--3\,GHz, as one would expect. These suggest that high-sensitivity multi-frequency data are necessary for a complete study of radio spectrum of SFGs. 

We also compare the measured radio spectral index at high and low frequencies for the 166 SFGs with measured flux densities at the three radio frequencies individually (Figure~\ref{f:figure3}). For the 76 SFGs that have $1<$\,SNR\,$\le3$ at GMRT 325\,MHz, we use the 5\,$\sigma$ limit of GMRT 325\,MHz data to estimate the lower limit of their $\alpha^{\rm 1.3GHz}_{\rm 325MHz}$. In addition, for the 90 SFGs that have SNR\,$>3$ at GMRT 325\,MHz, one of them has SNR\,$=2$ at MeerKAT 1.3\,GHz. Therefore, in total, there are 89 SFGs with SNR\,$>3$ at both MeerKAT 1.3\,GHz and GMRT 325\,MHz. We find that 33\% (29/89) of them have the difference between radio spectral indices at low and high frequencies ($\alpha^{\rm 1.3GHz}_{\rm 325MHz}-\alpha^{\rm 3.0GHz}_{\rm 1.3GHz}$) larger than the combined uncertainty, which is estimated as 
\begin{eqnarray} \label{e:equation2}
\sigma_{\rm combined}=\sqrt{(\sigma_{\alpha^{\rm 1.3GHz}_{\rm 325MHz}}^{2}+\sigma_{\alpha^{\rm 3.0GHz}_{\rm 1.3GHz}}^{2})}. 
\end{eqnarray}
In addition, one SFG has the estimated lower limit of ($\alpha^{\rm 1.3GHz}_{\rm 325MHz}-\alpha^{\rm 3.0GHz}_{\rm 1.3GHz}$) larger than its $\sigma_{\rm combined}$. We also mark the ten SFGs that have $\alpha^{\rm 3.0GHz}_{\rm 1.3GHz}<-1$ in Figure~\ref{f:figure3}. We find that five out of these ten have $z<0.5$, while four have $z=$\,0.7--0.8. Therefore, some of these sources' 3\,GHz flux densities are very likely affected by the missing flux of VLA observations. For SFGs brighter than 50\,$\mu$Jy at 3\,GHz, we thus estimate that $>15\%$ of them have a significantly flatter radio spectrum at low frequency.

%
% Table 1
%
\begin{table*}
\scriptsize
\caption{Results from stacking the 3\,GHz-selected SFGs}
\label{tab:table2}
\begin{tabular}{lcccc}
\hline
Redshift ($z$) & $z<0.5$ & 0.5--1.0 & 1.0--2.0 & $\ge$2.0 \\
\hline
Number & 373 & 668 & 703 & 350 \\
$S_{\rm 325MHz}$$^{*}$ & $203\pm9$ & $125\pm5$ & $109\pm4$  & $109\pm7$  \\
$S_{\rm 1.3GHz}$$^{*}$ & $99\pm4$ & $62\pm2$ & $56\pm1$ & $50\pm1$  \\
$\alpha^{\rm 1.3GHz}_{\rm 325MHz}$ & $-0.51\pm0.04$ & $-0.51\pm0.03$ & $-0.48\pm0.03$ & $-0.56\pm0.05$ \\
\end{tabular}
 
\begin{tabular}{lcc|ccc}
\hline
\multicolumn{3}{l|}{log$(M_{*}/\Msun)<$10.6} & \multicolumn{3}{l}{log$(M_{*}/\Msun)\ge$10.6} \\
\hline
$z<0.5$ & 0.5--1.0 & 1.0--2.0 & $z<1.0$ & 1.0--2.0 & $\ge$2.0 \\
\hline
 279 & 395 & 284 & 364 & 418 & 265\\
 $172\pm9$ & $106\pm4$ & $84\pm4$  & $191\pm10$ & $126\pm7$ & $121\pm9$ \\
 $88\pm4$ & $54\pm2$ & $46\pm1$ & $90\pm3$ & $63\pm2$ & $53\pm2$ \\
 $-0.49\pm0.05$ & $-0.49\pm0.04$ & $-0.44\pm0.04$ & $-0.55\pm0.04$ & $-0.50\pm0.05$ & $-0.60\pm0.06$  \\
 \hline
 \end{tabular}
 
 \begin{tabular}{lcccc}
\hline
 log$(M_{*}/\Msun)$ &  $<$10.3  & 10.3--10.6  & 10.6--10.9 & $\ge$10.9 \\
 \hline
 Number & 493 & 550 & 578 & 469 \\
 $S_{\rm 325MHz}$$^{*}$ & $112\pm5$ & $116\pm4$ & $141\pm6$  & $152\pm8$ \\
 $S_{\rm 1.3GHz}$$^{*}$ & $57\pm2$ & $62\pm2$ & $69\pm2$  & $70\pm2$ \\
$\alpha^{\rm 1.3GHz}_{\rm 325MHz}$ & $-0.49\pm0.03$ & $-0.46\pm0.03$ & $-0.52\pm0.04$ & $-0.56\pm0.04$ \\
\end{tabular}

\begin{tabular}{lccc|cccc}
\hline
 $z<$1  &  &  & & $z\ge1$  & &  &  \\
 \hline
  log$(M_{*}/\Msun)<$10.3  & 10.3--10.6  & 10.6--10.9 & $\ge$10.9  & log$(M_{*}/\Msun)<$10.3  & 10.3--10.6  & 10.6--10.9 & $\ge$10.9  \\
 \hline
 353 & 321 & 242 & 122 & 140 & 229 & 336 & 347 \\
 $125\pm6$ & $142\pm6$ & $188\pm13$  & $196\pm14$ & $78\pm6$ & $83\pm4$ & $111\pm6$ & $137\pm8$ \\
 $64\pm2$ & $72\pm3$ & $90\pm5$  & $90\pm4$ & $39\pm2$ & $47\pm2$ & $55\pm2$ & $63\pm2$ \\
$-0.48\pm0.04$ & $-0.49\pm0.05$ & $-0.54\pm0.06$ & $-0.55\pm0.05$ & $-0.51\pm0.06$ & $-0.41\pm0.05$ & $-0.51\pm0.04$ & $-0.56\pm0.05$ \\
\hline
\end{tabular}

\begin{tabular}{lcccc|cccc}
\hline
 SFR ($\Msun$\,yr$^{-1}$) & $z<$1  &   & & & $z\ge1$  & &  &  \\
 \hline
  & SFR$<$10  & 10--30 & 30--50  & $\ge$50  & SFR$<$100  & 100--200  & 200--400 & $\ge$400  \\
 \hline
Number &   239 & 348 & 251 & 203  & 254 & 301 & 297 & 201 \\
$S_{\rm 325MHz}$$^{*}$ &  $159\pm9$ & $146\pm7$ & $138\pm12$ & $176\pm8$  & $85\pm8$ & $90\pm4$ & $110\pm5$ & $167\pm13$ \\
 $S_{\rm 1.3GHz}$$^{*}$ & $79\pm4$ & $69\pm3$ & $65\pm4$ & $93\pm4$  & $38\pm2$ & $49\pm1$ & $59\pm2$ & $74\pm3$ \\
$\alpha^{\rm 1.3GHz}_{\rm 325MHz}$ &  $-0.50\pm0.05$ & $-0.54\pm0.04$ & $-0.55\pm0.08$ & $-0.46\pm0.04$ & $-0.58\pm0.07$ & $-0.44\pm0.04$ & $-0.46\pm0.04$ & $-0.58\pm0.06$ \\
\hline
\end{tabular}

\begin{tabular}{lcccc|cccc}
\hline
 sSFR & $z<$1  &   & & & $z\ge1$  & &  &  \\
\hline
& sSFR$<5.0\times10^{-10}$  & (0.5--1.0)$\times10^{-9}$ & (1.0--2.0)$\times10^{-9}$ & $\ge2.0\times10^{-9}$  & sSFR$<1.5\times10^{-9}$  & (1.5--3.0)$\times10^{-9}$  & (3.0--6.0)$\times10^{-9}$ & $\ge6.0\times10^{-9}$  \\
 \hline
Number &   331 & 278 & 245  & 184 & 194 & 282 & 288 & 288 \\
$S_{\rm 325MHz}$$^{*}$ &  $172\pm9$ & $157\pm12$ & $137\pm6$  & $135\pm9$ & $117\pm11$ & $89\pm6$ & $120\pm7$ & $112\pm6$ \\
 $S_{\rm 1.3GHz}$$^{*}$ & $82\pm4$ & $76\pm4$ & $69\pm3$  & $69\pm4$ & $51\pm2$ & $47\pm2$ & $56\pm2$ & $59\pm2$ \\
$\alpha^{\rm 1.3GHz}_{\rm 325MHz}$ &  $-0.53\pm0.05$ & $-0.52\pm0.07$ & $-0.49\pm0.04$ & $-0.48\pm0.06$ & $-0.59\pm0.07$ & $-0.46\pm0.05$ & $-0.54\pm0.05$ & $-0.46\pm0.05$ \\
\hline
\end{tabular}

 \begin{tabular}{lccccc}
\hline
 $\alpha^{\rm 3GHz}_{\rm 1.3GHz}$ &  $<-1.0$  & $(-1.0)$--$(-0.8)$  & $(-0.8)$--$(-0.65)$ & $(-0.65)$--$(-0.4)$ & $\ge-0.4$ \\
 \hline
 Number & 469 & 472 & 396 & 413 & 262 \\
 $S_{\rm 325MHz}$$^{*}$ & $129\pm4$ & $139\pm5$ & $138\pm5$  & $114\pm5$ & $65\pm4$ \\
 $S_{\rm 1.3GHz}$$^{*}$ & $61\pm2$ & $68\pm2$ & $71\pm2$  & $59\pm2$ & $37\pm1$ \\
$\alpha^{\rm 1.3GHz}_{\rm 325MHz}$ & $-0.54\pm0.03$ & $-0.52\pm0.03$ & $-0.48\pm0.03$ & $-0.48\pm0.04$ & $-0.40\pm0.05$ \\
\end{tabular}

\begin{tabular}{lcccc|ccccc}
\hline
 $z<$1  &  &  & & & $z\ge1$  & &  & & \\
 \hline
  $\alpha^{\rm 3GHz}_{\rm 1.3GHz}<-1.0$  & $(-1.0)$--$(-0.8)$  & $(-0.8)$--$(-0.65)$ & $(-0.65)$--$(-0.4)$ & $\ge-0.4$  & $\alpha^{\rm 3GHz}_{\rm 1.3GHz}<-1.0$  & $(-1.0)$--$(-0.8)$  & $(-0.8)$--$(-0.65)$ & $(-0.65)$--$(-0.4)$ & $\ge-0.4$ \\
 \hline
 258 & 233 & 215 & 186 & 106 & 211 & 239 & 181 & 227 & 156 \\
 $137\pm6$ & $159\pm8$ & $160\pm8$  & $125\pm8$ & $73\pm7$ & $118\pm7$ & $120\pm6$ & $113\pm6$ & $105\pm6$ & $59\pm5$ \\
 $65\pm2$ & $77\pm3$ & $81\pm3$  & $65\pm3$ & $41\pm2$ & $56\pm2$ & $59\pm2$ & $60\pm2$ & $53\pm2$ & $34\pm2$ \\
$-0.54\pm0.04$ & $-0.52\pm0.04$ & $-0.50\pm0.05$ & $-0.47\pm0.06$ & $-0.42\pm0.07$ & $-0.54\pm0.05$ & $-0.51\pm0.04$ & $-0.45\pm0.05$ & $-0.49\pm0.05$ & $-0.39\pm0.07$ \\
\hline
\end{tabular}
\\
$^*$Flux density in $\mu$Jy.
\end{table*}

\subsubsection{Stacking all of the 3\,GHz-selected SFGs}\label{s:lower limit}

For all of the 2,094 VLA 3\,GHz selected SFGs in our sample, only 195 of them have SNR\,$>3$ at GMRT 325\,MHz. In addition, the cut of SNR\,$>3$ at GMRT 325\,MHz biases our sample towards the SFGs with relatively steeper radio spectrum observed at 0.33--3\,GHz as described in Section $\S$ \ref{s:individual measure}. We therefore perform a stacking analysis to study the radio spectral properties at observer-frame 0.33--1.3\,GHz.

For these 2,094 SFGs, we stack both GMRT 325\,MHz and MeerKAT 1.3\,GHz imaging data at their 3\,GHz positions to measure the averaged radio spectral index at observer-frame 0.33--1.3\,GHz. We use the peak flux densities of the average-stacked images to estimate the radio spectral index at low frequency and show the results in Figure~\ref{f:radio_sindex_snr5_mightee_3GHz} and Table~\ref{tab:table2}. To compare the correlations of radio spectral index and physical properties of galaxies shown in Figure~\ref{f:rest_3_z}, we stack the 3\,GHz-selected SFGs by redshift, stellar mass, SFR, and sSFR bins respectively. 

As shown in Figure~\ref{f:radio_sindex_snr5_mightee_3GHz} and Table~\ref{tab:table2}, when stacking the SFGs within different redshift bins, we further divide the sample into massive (log$(M_{*}/\Msun)\ge10.6$) and less-massive (log$(M_{*}/\Msun) <10.6$) subsamples. {We find} that the averaged radio spectral slope at low-frequency slightly steepens at $\nu_{\rm rest, 325MHz}\ga1.0$\,GHz (plot a of Figure~\ref{f:radio_sindex_snr5_mightee_3GHz}). However, as shown in Figure~\ref{f:radio_sindex_snr5_mightee_3GHz}, our sample at $z>2$ are dominated by massive SFGs. Therefore, the results suggest that, on average, massive SFGs have a relatively steeper radio spectrum at observer-frame 0.33--1.3\,GHz.

For the 2,094 SFGs, 2,090 of them have estimated stellar mass from the ``super-deblended" catalogue \citep{Jin18}. We further divide them into high-redshift ($z\ge1$) and low-redshift ($z<1$) subsets to investigate the selection effect of our radio flux-limited sample. Except the lowest stellar mass bin, which is affected the most by the incompleteness of our flux-limited sample, the results from stacking show that the averaged radio spectral slope at low frequency also steepens with increasing stellar mass (Figure~\ref{f:radio_sindex_snr5_mightee_3GHz}).

We also divide our sample into high-redshift ($z\ge1$) and low-redshift ($z<1$) subsets when stacking SFGs within different SFR bins and confirm that the trend between the averaged radio spectral index and SFR is caused by the redshift-dependent selection effect of our radio flux-limited sample (plot c of Figure~\ref{f:radio_sindex_snr5_mightee_3GHz}). As shown in the plot  d of Figure~\ref{f:radio_sindex_snr5_mightee_3GHz}, the averaged radio spectral slope at low frequency also flattens with increasing sSFR in general, although the high-redshift subsample exhibits the greater fluctuation in the trend.

For the 2,012 SFGs with measured radio spectral index at high frequency ($\alpha^{\rm 3GHz}_{\rm 1.3GHz}$), we stack them by high-frequency radio spectral index bins and show the results in Table~\ref{tab:table2}. We find that the averaged radio spectral index at low frequency is positively correlated with that at high frequency, i.e., the SFGs within the lowest $\alpha^{\rm 3GHz}_{\rm 1.3GHz}$ bin also have the steepest averaged radio spectrum at low frequency, and vice versa. We further divide the 2,012 SFGs into high-redshift ($z\ge1$) and low-redshift ($z<1$) subsamples and find that the positive correlation between the averaged radio spectral indices at low and high frequencies is present in both redshift subsamples despite some fluctuations.

Overall, the stacking analyses of all the 3\,GHz-selected SFGs suggest that the trends between radio spectral index and physical properties of SFGs are consistent at low and high frequencies. In addition, the averaged radio spectral index at low and high frequencies show a positive correlation. 

%
% Figure 5
%
\begin{figure}
\centering
\includegraphics[width=0.48\textwidth]{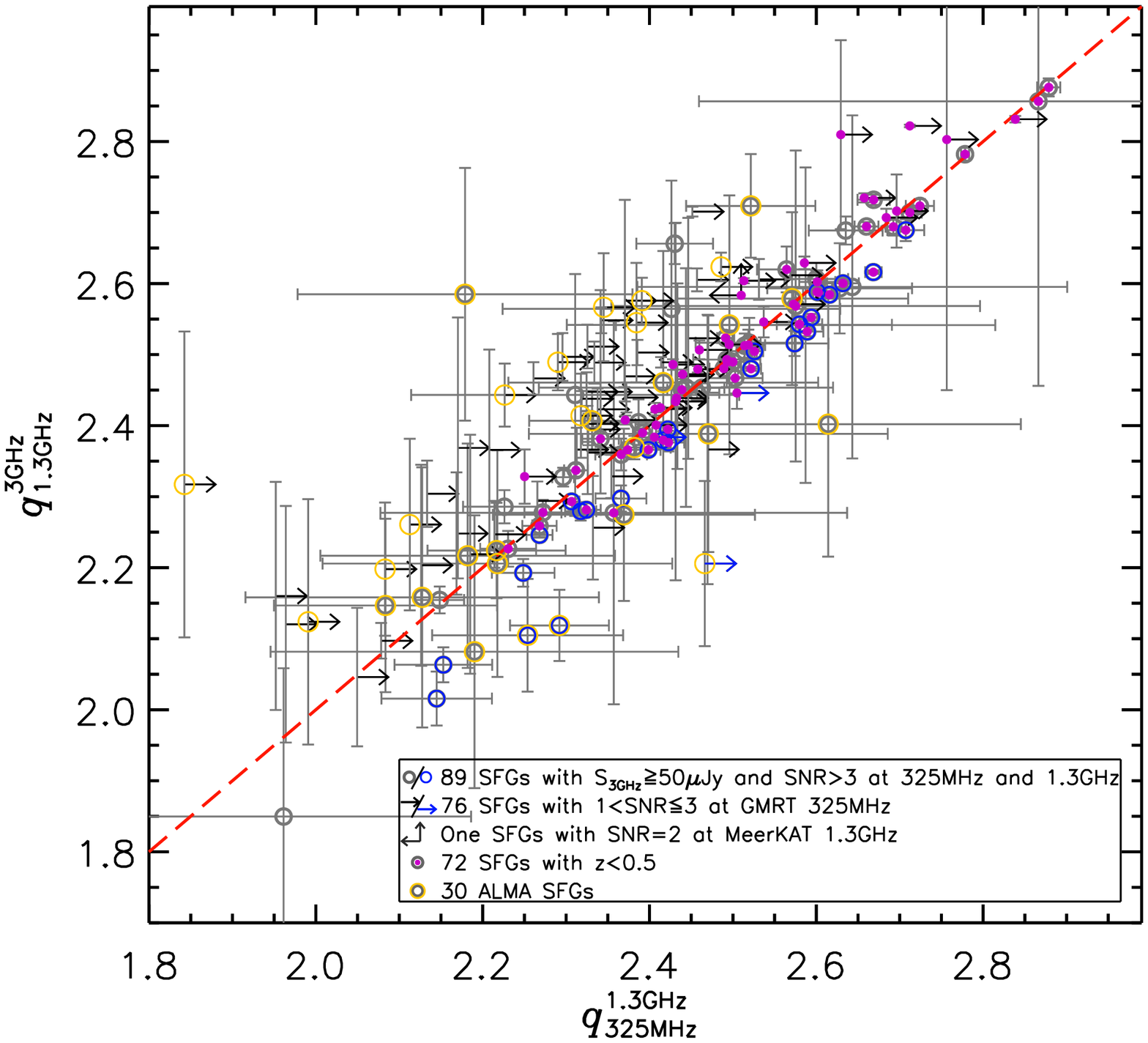}
\caption{Comparison of $q_{\rm IR}$ based on radio spectral indices at low and high frequencies for the 166 SFGs with $S_{\rm 3GHz}\ge50\,\mu$Jy and measured flux densities at the three radio frequencies. The $q_{\rm 325MHz}^{\rm 1.3GHz}$ is measured by adopting $\alpha_{\rm 325MHz}^{\rm 1.3GHz}$ to $k$-correct the observed flux density, while $q_{\rm 1.3GHz}^{\rm 3GHz}$ is based on $\alpha_{\rm 1.3GHz}^{\rm 3GHz}$-derived $k$-correction. The circles are the 89 SFGs with SNR\,$>3$ at both GMRT 325\,MHz and MeerKAT 1.3\,GHz. For the 76 SFGs with $1<$\,SNR\,$\le3$ at GMRT 325\,MHz, we use the lower limit of their radio spectral indices to estimate the lower limit of their $q_{\rm 325MHz}^{\rm 1.3GHz}$ as shown by rightward arrows. The remaining one SFG has SNR\,$>3$ at GMRT 325\,MHz but has SNR\,$=2$ at MeerKAT 1.3\,GHz. We show its upper limit of $q_{\rm 325MHz}^{\rm 1.3GHz}$ and lower limit of $q_{\rm 1.3GHz}^{\rm 3GHz}$ by leftward and upward arrows. 
We also mark the SFGs with the measured ($q_{\rm 325MHz}^{\rm 1.3GHz}-q_{\rm 1.3GHz}^{\rm 3GHz}$) larger than their combined uncertainties by blue symbols. Errors of $q_{\rm IR}$ are derived from the uncertainties of infrared and radio luminosities.
}
\label{f:fir_radio_correlation_q}
\end{figure}

%
% Figure 6
%
\begin{figure*}
\centering
\includegraphics[width=0.96\textwidth]{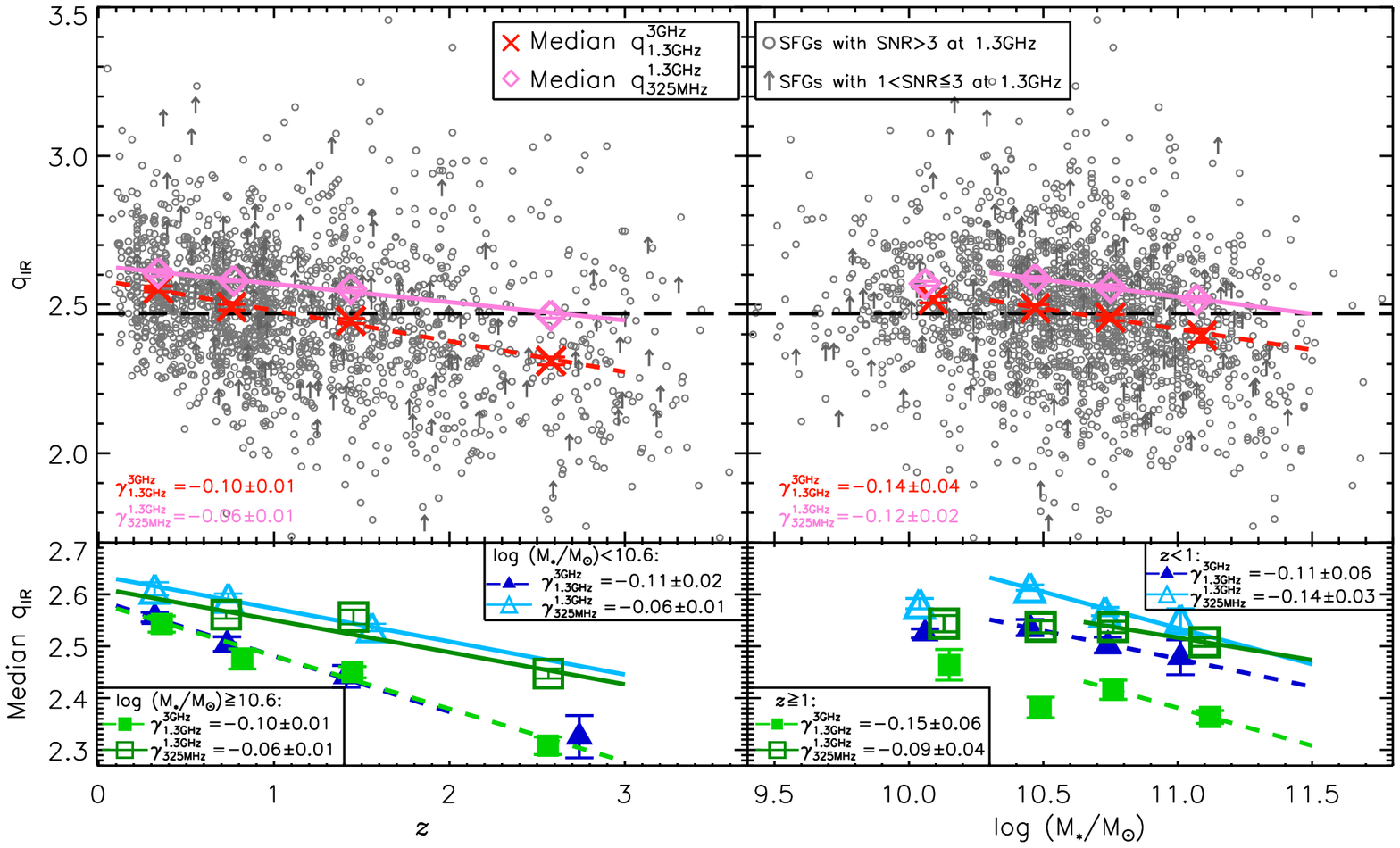}
\caption{{\it Top}: Logarithmic infrared-to-radio luminosity ratio, $q_{\rm IR}$ as functions of redshift and stellar mass for the 1,835 VLA 3\,GHz-selected SFGs with measured spectral index $\alpha_{\rm 1.3GHz}^{\rm 3GHz}$ and SNR$_{\rm IR}>3$. Among them, 1,717 also have SNR\,$>3$ at MeerKAT 1.3\,GHz. Their $q_{\rm IR}$ are measured by adopting $\alpha_{\rm 1.3GHz}^{\rm 3GHz}$ in $k$-correction. The black dashed line represents the median $q_{\rm IR}$ ($q_{\rm IR}=2.48\pm0.01$) of these SFGs. The red crosses show the median $q_{\rm IR}$ based on $\alpha_{\rm 1.3GHz}^{\rm 3GHz}$ of galaxies within each redshift and stellar mass bin while the error bars show the errors that are estimated from bootstrap resampling. For the remaining 118 SFGs with $1<$\,SNR\,$\le3$ at MeerKAT 1.3\,GHz, we use the lower limit of their $\alpha_{\rm 1.3GHz}^{\rm 3GHz}$ to estimate the lower limit of their $q_{\rm IR}$ as shown by upward arrows. We also adopt the averaged radio spectral index at low frequency ($\alpha_{\rm 325MHz}^{\rm 1.3GHz}$) from our stacking analyses (Table~\ref{tab:table2}) for galaxies within each bin and show the median $q_{\rm 325MHz}^{\rm 1.3GHz}$ as magenta diamond. The solid and short-dashed lines represent the linear fits of correlations between median $q_{\rm IR}$ and physical properties based on the two different $k$-corrections respectively. The slopes of these linear fits ($\gamma$) are shown on the bottom-left corner of each panel. We do not include the lowest stellar mass bin in the fits to reduce the effect of incompleteness of our flux-limited sample. {\it Bottom}: We further divide our sample into massive (log $(M_{*}/\Msun)\ge10.6$) and less-massive (log $(M_{*}/\Msun)<10.6$), high-redshift ($z\ge1$) and low-redshift ($z<1$) subsamples and show their median $q_{\rm IR}$ as functions of redshift and stellar mass. The solid symbols represent the median $q_{\rm 1.3GHz}^{\rm 3GHz}$ while the open symbols show the median $q_{\rm 325MHz}^{\rm 1.3GHz}$, which is estimated by adopting the averaged $\alpha_{\rm 325MHz}^{\rm 1.3GHz}$ from our stacking analyses. The highest redshift bin of less massive subsample, the lowest stellar mass bin of low-redshift subset, and the two lowest stellar mass bins of high-redshift subsample are excluded in  the linear fits to reduce the effect of incompleteness of our radio flux-limited sample. Overall, the two different $k$-corrections result in a similar overall trend between $q_{\rm IR}$ and redshift or stellar mass, although evolution of $q_{\rm IR}$ with cosmic time will be slightly overestimated if we adopt the radio spectral index from high frequency ($\alpha_{\rm 1.3GHz}^{\rm 3GHz}$) in $k$-correction.
}
\label{f:frc_evolution}
\end{figure*}

\subsection{FIR-Radio correlation}\label{s:FIR-radio}
With the measured radio spectral indices of SFGs, we study the correlation between the infrared and radio emission of SFGs and investigate how different assumptions of $k$-correction affect this. In this work we use rest-frame radio luminosities at 1.3\,GHz ($L_{\rm 1.3GHz}$) in the FIRRC of SFGs to be consistent with the observational frequency of our MIGHTEE survey. 

\subsubsection{FIR-Radio correlation for bright 3\,GHz SFGs}\label{s:FIR-radio-bright}

To measure $L_{\rm 1.3GHz}$, the spectral index at low frequency ($\alpha^{\rm 1.3GHz}_{\rm 325MHz}$) is essential because the emission at rest-frame 1.3\,GHz from a galaxy with $z>0$ is shifted into $\nu_{\rm obs}<$\,1.3\,GHz. Therefore, the appropriate spectral index to use in $k$-correction should be $\alpha^{\rm 1.3GHz}_{\rm 325MHz}$. However, as we described in Section $\S$\ref{s:sed_super}, because of the low sensitivity of the GMRT data, only the sample of the SFGs brighter than 50\,$\mu$Jy at 3\,GHz has a relatively complete ($>$\,80\%) measurement of flux densities at GMRT 325\,MHz. We thus first use $\alpha^{\rm 1.3GHz}_{\rm 325MHz}$ to $k$-correct the observed flux density at MeerKAT 1.3\,GHz and derive the rest-frame luminosity at 1.3\,GHz for the 89 SFGs with SNR\,$>3$ at both GMRT 325\,MHz and MeerKAT 1.3\,GHz. For the 76 SFGs with $1<$\,SNR\,$\le3$ at GMRT 325\,MHz, we use the lower limit of their $\alpha^{\rm 1.3GHz}_{\rm 325MHz}$. The remaining one SFG has SNR\,$>3$ at GMRT 325\,MHz but has SNR\,$=2$ at MeerKAT 1.3\,GHz. We therefore use the lower limit of its $\alpha^{\rm 1.3GHz}_{\rm 325MHz}$ to estimate the upper limit of $q^{\rm 1.3GHz}_{\rm 325MHz}$ (Figure~\ref{f:fir_radio_correlation_q}). Combining with their total infrared luminosities from the ``super-deblended'' catalogue \citep{Jin18}, we measure the logarithmic infrared-to-radio luminosity ratio \citep[e.g.,][]{Helou85, Condon92, Ivison10}: 
\begin{eqnarray} \label{e:equation3}
q_{\rm IR}={\rm log}[\frac{L_{\rm IR}}{(3.75\times10^{12}\,W)}\times \frac{\rm W Hz^{-1}}{L_{\rm 1.3\,GHz}}].
\end{eqnarray}

We obtain a median  $q_{\rm IR}=2.46^{+0.03}_{-0.04}$ and a scatter of $\sigma=0.18$ for the 89 SFGs with SNR\,$>3$ at both GMRT 325\,MHz and MeerKAT 1.3\,GHz. To investigate how different $k$-corrections affect the FIRRC of SFGs, we also use the spectral indices at high frequency ($\alpha^{\rm 3GHz}_{\rm 1.3GHz}$) to measure FIRRC and obtain a median $q_{\rm IR}=2.46^{+0.03}_{-0.02}$ and a standard deviation of $\sigma=0.19$. In addition, we assume a fixed radio spectral slope with $\alpha=-0.8$ at rest-frame 0.4--10\,GHz to $k$-correct the observed flux density at 1.3\,GHz and obtain a median $q_{\rm IR}=2.45^{+0.03}_{-0.04}$ and a scatter of $\sigma=0.18$. Therefore, for these SFGs with $S_{\rm 3GHz} \ge50\,\mu$Jy and SNR\,$>3$ at GMRT 325\,MHz and MeerKAT 1.3\,GHz, the adoption of a fixed or an individually measured radio spectral index at high frequency does not significantly affect the median $q_{\rm IR}$. However, as described in Section $\S$\ref{s:individual measure}, the cut of SNR\,$>3$ at GMRT 325\,MHz biases our sample towards the SFGs with a relatively steeper radio spectrum at observer-frame 0.33--3\,GHz. This might be the reason that our median $q_{\rm IR}$ is slightly lower than the measurements of some previous works \citep[e.g.,][]{Bell03, Ivison10, Thomson14, Molnar21}.

We show the $q_{\rm IR}$ based on radio spectral indices at low and high frequencies for these 166 SFGs in Figure~\ref{f:fir_radio_correlation_q}. For the 89 SFGs with SNR\,$>3$ at both GMRT 325\,MHz and MeerKAT 1.3\,GHz, we find that 28\% (25/89) of them have the measured ($q_{\rm 325MHz}^{\rm 1.3GHz}-q_{\rm 1.3GHz}^{\rm 3GHz}$) larger than their combined uncertainties. Here the ``combined uncertainty'' is defined in an analogous way to the $\sigma_{\rm combined}$ of Equation~\ref{e:equation2}, using the uncertainties of $q_{\rm 325MHz}^{\rm 1.3GHz}$ and $q_{\rm 1.3GHz}^{\rm 3GHz}$. In addition, three SFGs has the estimated lower limit of ($q_{\rm 325MHz}^{\rm 1.3GHz}-q_{\rm 1.3GHz}^{\rm 3GHz}$) larger than their combined uncertainties. Therefore, for the SFGs brighter than 50\,$\mu$Jy at 3\,GHz, $>$\,17\% (28/166) of their $q_{\rm IR}$ are significantly underestimated if we adopt the radio spectral index at high frequency in the $k$-correction.

\subsubsection{FIR-Radio correlation for all of the 3\,GHz-selected SFGs}\label{s:FIR-radio-all}

Unfortunately, the majority of the 3\,GHz-selected SFGs in our sample lack the measured flux density at GMRT 325\,MHz because of the low sensitivity of 325\,MHz data. We thus first use $\alpha_{\rm 1.3GHz}^{\rm 3GHz}$ instead of $\alpha_{\rm 325MHz}^{\rm 1.3GHz}$ to $k$-correct the observed luminosity at MeerKAT 1.3\,GHz and measure the FIRRC for the 2,012 VLA 3\,GHz-selected SFGs (Figure~\ref{f:frc_evolution}). Among them, 161 have $1<$\,SNR\,$\le3$ at MeerKAT 1.3\,GHz. We therefore use the lower limit of their $\alpha_{\rm 1.3GHz}^{\rm 3GHz}$ to estimate the lower limit of their $q_{\rm IR}$ as shown in Figure~\ref{f:frc_evolution}. In addition, for the 2,012 radio-selected SFGs, 1,835 of them have SNR$_{\rm IR}>3$. The SNR$_{\rm IR}$ is the combined SNR of ``super-deblended'' flux densities from 100\,$\mu$m-to-1.2\,mm, which reflects the quality of the FIR to millimeter SEDs that are used in estimating the total infrared luminosity. We therefore only keep the 1,835 SFGs with SNR$_{\rm IR}>3$ in the following analyses. Among them, 1,717 also have SNR\,$>3$ at MeerKAT 1.3\,GHz. Their median $q_{\rm IR}=2.47\pm0.01$, with a standard deviation of $\sigma=0.23$, is consistent with that of SFGs brighter than 50\,$\mu$Jy at 3\,GHz (Section $\S$\ref{s:FIR-radio-bright}).

As shown in the top panels of Figure~\ref{f:frc_evolution}, there is a clear trend between $q_{\rm IR}$ and redshift/stellar mass of galaxies, namely, that $q_{\rm IR}$ slightly declines with redshift and stellar mass. The evolution of $q_{\rm IR}$ with redshift and/or stellar mass has been widely discussed in literature \citep[e.g.,][]{Magnelli15, Delhaize17, Jarvis10, Algera20, Delvecchio21}. However, a complete study of the evolution of $q_{\rm IR}$ and its physical factors are beyond the scope of this work. As described in Section $\S$\ref{s:super-deblended_SFGs}, we remove radio excess sources from our sample to select a clean sample of SFG, which affects the completeness of study FIRRC of radio-selected sources. In this work, we only focus on how different $k$-corrections affect the study of FIRRC of SFGs.

To compare the FIRRC based on radio spectral index at low and high frequencies for all the 3\,GHz-selected SFGs, we first divide our sample into the same redshift and stellar mass bins as those listed in Table~\ref{tab:table2} and show the median $q_{\rm 1.3GHz}^{\rm 3GHz}$ of the galaxies within each bin in Figure~\ref{f:frc_evolution}. We also adopt the radio spectral index at low frequency, $\alpha_{\rm 325MHz}^{\rm 1.3GHz}$, from our stacking analyses (Table~\ref{tab:table2}) to $k$-correct the observed flux density at MeerKAT 1.3\,GHz of galaxies within each bin and show the median $q_{\rm 325MHz}^{\rm 1.3GHz}$ in Figure~\ref{f:frc_evolution}. We first perform the simple linear fit to the correlations between median $q_{\rm IR}$ and physical properties of galaxies within each bin and overplot the results in Figure~\ref{f:frc_evolution}. To be consistent with previous studies, we also fit the evolution of median $q_{\rm IR}$ as a function of redshift with the form $q\propto(1+z)^{\gamma}$ \citep[e.g.,][]{Ivison10, Delvecchio21} and obtain the best-fit $\gamma=-0.10\pm0.01$ for $q_{\rm 1.3GHz}^{\rm 3GHz}$ and $\gamma=-0.05\pm0.01$ for $q_{\rm 325MHz}^{\rm 1.3GHz}$. We further divide our sample into massive (log $(M_{*}/\Msun)\ge10.6$) and less massive (log $(M_{*}/\Msun)<10.6$) subsamples and show the median $q_{\rm 1.3GHz}^{\rm 3GHz}$ and $q_{\rm 325MHz}^{\rm 1.3GHz}$ as a function of redshift in the bottom-left panel of Figure~\ref{f:frc_evolution}. We exclude the highest redshift bin of the less massive galaxies in the fit to reduce the effect of incompleteness in our sample. Overall, Figure~\ref{f:frc_evolution} and our analyses show that although the two different $k$-corrections result in a similar overall trend between $q_{\rm IR}$ and redshift, the evolution of $q_{\rm IR}$ with cosmic time based on radio spectral index at low frequency is slightly weaker than that based on spectral index derived from high frequency. 

The bottom-right plot of Figure~\ref{f:frc_evolution} shows the median $q_{\rm IR}$ as a function of stellar mass for SFGs at high redshift ($z\ge1$) and low redshift ($z<1$) respectively. To reduce the effect from incompleteness of our radio flux-limited sample, we do not include the lowest stellar mass bin and the two lowest stellar mass bins in fitting the correlations for the low-redshift and high-redshift subsamples respectively. As shown in Figure~\ref{f:frc_evolution}, the trends between  $q_{\rm IR}$ and stellar mass based on the two radio spectral indices are consistent within the uncertainties. 

Overall, using the radio spectral index from high frequency or the fixed spectral index of $\alpha=-0.8$ to $k$-correct the observed radio flux density will underestimate the $q_{\rm IR}$ for galaxies with a flatter radio spectrum at low frequency, and thus further affect the slope of the evolution of $q_{\rm IR}$ with redshift.

\section{Discussion}\label{s:discussion}
In this work, we have presented the radio spectral properties observed at 0.33--1.3\,GHz and 1.3--3\,GHz respectively for the 2,094 VLA 3\,GHz-selected SFGs. Here we discuss the possible physical mechanisms that determine the radio spectrum at low and high frequencies.

%%%%%

\subsection{Physical mechanisms that affect the radio spectrum at high frequency}\label{s:discussion_high}

In Section $\S$\ref{s:results_super}, we show that, for the 2,012 SFGs with $z\sim$\,0.01--3, the median radio spectral index between the observer-frame frequencies of 1.3--3\,GHz is $\alpha_{\rm 1.3GHz}^{\rm 3GHz}=-0.78\pm0.01$. In addition, the median radio spectral index does not change significantly at rest-frame frequencies of 1.3--10\,GHz, which suggests that the radio spectrum of SFGs at this frequency range is dominated by the synchrotron emission.

Figure~\ref{f:rest_3_z} shows that, on average, the radio spectrum at observer-frame 1.3--3\,GHz slightly steepens with increasing stellar mass. One possible explanation for this correlation is the ``aging'' of the relativistic electron population. In the literature, the CR electron population losing energy with time, i.e., via inverse-Compton, synchrotron, ionization, and bremstrahlung process, is widely used to explain a steepening of the radio spectrum at high frequency \citep[e.g.,][]{Condon92, Basu15, Klein18, Chyzy18, Thomson19}. However, for SFGs, the constantly injected CR electrons in star-forming regions will keep the radio spectral slope constant. Therefore, the slope of radio spectrum relies on the balance between the freshly injected and aged CR electrons in SFGs \citep[e.g,][]{Basu15, Tabatabaei17}. The correlation between radio spectral index and stellar mass shown in Figure~\ref{f:rest_3_z} and Figure~\ref{f:radio_sindex_snr5_mightee_3GHz} could be explained by slightly increased ratio of aged and young relativistic CR electrons with increasing stellar mass of SFGs, although the data and the analyses in this work are not sufficient for us to be completely confident in this scenario. 

We also find a weak positive correlation between sSFR and the radio spectral index at observer-frame 1.3--3\,GHz of SFGs. We notice that the radio spectrum flattening with increasing sSFR has also been observed in local ULIRGs \citep{Condon91, Murphy13}. However, our results shown in Figure~\ref{f:rest_3_z} suggest that this trend in our sample might be an alternative manifestation of the underlying physics as indicated by the correlation between the radio spectrum and the stellar mass of SFGs. \cite{Murphy13} suggested that galaxies with high sSFR are more compact and host deeply embedded star formation, thus more optically thick in the radio. Therefore, the radio spectrum flattening with increasing sSFR could be explained by increased free-free absorption \citep{Murphy13}. However, further studies are necessary to explore the underlying physics that drive the correlations between radio spectral index and stellar mass, and sSFR of SFGs.

\subsection{Physical mechanisms that flatten the radio spectrum at low frequency}\label{s:discussion_low}

Our analyses for all of the 3\,GHz-selected SFGs and the subset brighter than 50\,$\mu$Jy at 3\,GHz both show that, on average, the radio spectrum of SFGs flattens at low frequency ($\nu_{\rm obs}<1.3$\,GHz). Some recent works on the radio spectral properties of SFGs have also reported a flatter radio spectrum at low frequency \citep[e.g.,][]{Calistro17, Tisanic19,  Thomson19}, although their sample selections and/or observer-frame frequencies are different from this work.

There are several physical mechanisms that might explain a flattening of the radio spectrum at low frequency, such as thermal absorption, the intrinsic curvature in the synchrotron spectrum due to energy losses of CR electrons and propagation effects, Razin effect, relativistic bremsstrahlung, synchrotron self-absorption, and so on \citep[e.g.,][]{Condon92,McDonald02, Tingay04, Murphy09, Clemens10, Lacki13, Marvil15, Kapinska17, Chyzy18, Klein18}. Among them, the thermal absorption and the intrinsic curvature in the synchrotron spectrum are the two main mechanisms that are widely included in the theoretical models in the literature. Models of the thermal free-free absorption assume that a power-law synchrotron radio spectrum is attenuated by free-free absorption with an absorption coefficient: 
\begin{eqnarray} \label{e:equation4}
\left(\frac{\kappa}{\rm pc^{-1}}\right) \approx 3.3\times10^{-7}\left(\frac{n_{e}}{\rm cm^{-3}}\right)^2\left(\frac{T_{e}}{\rm 10^{4}\,K}\right)^{-1.35}\left(\frac{\nu}{\rm GHz}\right)^{-2.1},
\end{eqnarray}
where $n_{\rm e}$ is the free electron density, $T_{\rm e}$ is the electron temperature, and $\nu$ is the radio frequency \citep{Condon92}. 

Alternatively, models based on the intrinsic curvature in the synchrotron spectrum usually assume that the shock-accelerated CRs in the supernova remnants (SNRs) have a power-law energy distribution with spectral index $\alpha\sim-0.5$, which is supported by the averaged radio spectral index at $\sim$1\,GHz of SNRs \citep[e.g.,][]{Green14, Marvil15,Klein18,Chyzy18}. As the relativistic CR electrons propagate within the galaxy, two major mechanisms cause energy-loss at high frequency, namely synchrotron and inverse-Compton losses, which steepens the radio spectrum of SFGs. Although the remaining mechanisms may also flatten the radio spectrum at low frequency, previous work suggested that their impact in SFGs are very weak at rest-frame $\sim$\,0.4--1\,GHz \citep[e.g.,][]{Tingay04,Lacki13,Chyzy18}.

In our work, as shown in Figure~\ref{f:rest_3_z}, Figure~\ref{f:radio_sindex_snr5_mightee_3GHz}, and Table~\ref{tab:table2}, on average, the radio spectrum at both low and high frequency slightly steepens with increasing stellar mass, which could be explained by age-related synchrotron losses. These results lend support to the models that take into account energy-loss mechanisms in explaining the flattening of the radio spectrum at low frequency. However, some recent studies based on nearby bright galaxies suggest that the different regions within a galaxy might have distinctly different radio spectra because of the inhomogeneous physical conditions within a galaxy \citep[e.g.,][]{Marvil15, Kapinska17}. Therefore, models with a single mechanism may not be adequate for interpreting the integrated radio spectrum of a galaxy.

\subsection{Limitations}

As described in Section $\S$\ref{s:Analysis}, using 3\,GHz as the selection frequency in this work provides a less-biased sample for studying radio spectrum at observer-frame 1.3--3\,GHz. However, there might be some 1.3\,GHz sources with steep radio spectra, which would not have been included in this work because of the non-detection at 3\,GHz.

For studying the radio spectrum at observer-frame 0.33--3\,GHz, the main limitation is the small sample size of SFGs with measured radio spectral index at both low and high frequency because of the low sensitivity of GMRT 325\,MHz data. Although we applied a higher SNR cut and an additional flux density cut at 3\,GHz to provide a less-biased sample when studying the radio spectral properties of SFGs, nearly half (46\%) of SFGs with a measured radio spectral index at low frequency ($\alpha_{\rm 325MHz}^{\rm 1.3\,GHz}$) have a SNR\,$<$ 3 at 325\,MHz. This leads to large uncertainties on the measured radio spectral index at low frequency and reduces the accuracy of the follow-up analyses. On the other hand, the cut of SNR\,$>$ 3 at 325\,MHz bias our sample towards the SFGs with relatively steeper radio spectra at observer-frame 0.33-3\,GHz.

The other limitation is the uncertainties of the ``super-deblended'' flux densities extracted from imaging data with very different angular resolutions. As shown in Figure~\ref{f:check_vla_mightee_flux}, although the  ``super-deblended'' flux densities extracted from VLA 1.4\,GHz and MeerKAT 1.3\,GHz imaging data are consistent with each other within the uncertainties, the scatter is large. To improve quality and sample size in studying the radio spectral properties at both low and high frequency, observations with similar spatial resolutions and high sensitivity at multiple frequencies are essential. The recently released high-sensitivity low-frequency ($<$\,150\,MHz) data from the Low Frequency Array (LOFAR) large surveys \citep[e.g.,][]{de21, Tasse21, Sabater21} and the undergoing joint project, SuperMIGHTEE, which combines the MeerKAT and uGMRT telescopes to observe the MIGHTEE survey regions at 0.5--2.7\,GHz with the same angular resolution ($\sim5\arcsec$) and similar sensitivity, will dramatically advance the studies of the radio spectrum.

\section{Conclusion}\label{s:conclusion}
By combining deep MeerKAT 1.3\,GHz data with VLA 3\,GHz, GMRT 325\,MHz data and rich ancillary data available in the COSMOS field, we study the radio spectral properties of 2,094 radio 3\,GHz-selected SFGs within the MIGHTEE-COSMOS early science coverage. Our main conclusions are as follows.

1. After removing AGN and red quiescent galaxies from the 4,014 VLA 3\,GHz-selected sources within the MIGHTEE-COSMOS early science coverage, we obtain 2,094 SFGs with SNR\,$>$\,5 at 3\,GHz. We extract flux densities from the MeerKAT imaging data for these 3\,GHz-selected SFGs and obtain the radio spectral indices between the observer-frame frequencies of 1.3--3\,GHz for 96\% (2,012/2,094) of them. We limit our sample to the SFGs with SNR\,$>$3 at MeerKAT 1.3\,GHz and obtain a median spectral index of $\alpha_{\rm 1.3GHz}^{\rm 3GHz}=-0.80\pm0.01$. Although the scatter is large, the median radio spectral slope of these SFGs does not change within rest-frame $\sim$1.3--10\,GHz. This confirms that on average the radio spectrum of SFGs at rest-frame $\sim$1.3--10\,GHz is dominated by synchrotron emission. 

2. On average, the radio spectrum at observer-frame 1.3--3\,GHz slightly steepens with increasing stellar mass, and flattens with sSFR of SFGs, which may reflect the same fundamental relation. These trends could be explained by age-related synchrotron losses, i.e., the CR electron population losing energy with time, steepening the radio spectrum at rest-frame $\sim$1.3--10\,GHz of SFGs.

3. We extract flux densities from the GMRT 325\,MHz data for the 3\,GHz-selected SFGs in our sample. Due to the lower sensitivity of GMRT 325\,MHz data, we apply a further flux density cut ($S_{\rm 3GHz}\ge50\,\mu$Jy) to obtain a less-biased sample to study the radio spectral properties at observer-frame 0.33--3\,GHz. The median radio spectral indices for the 166 SFGs with measured flux densities in the three radio frequencies are $\alpha_{\rm 325MHz}^{\rm 1.3\,GHz}=-0.59^{+0.02}_{-0.03}$ and $\alpha_{\rm 1.3GHz}^{\rm 3\,GHz}=-0.74^{+0.01}_{-0.02}$. Therefore, on average, the radio spectrum of SFGs flattens at low frequency (0.33--1.3\,GHz).

4. For the SFGs that lack measured flux densities at GMRT 325\,MHz, we stack both GMRT 325\,MHz and MeerKAT 1.3\,GHz imaging data at their 3\,GHz positions and estimate their averaged radio spectral index at observer-frame 0.33--1.3\,GHz within each redshift, stellar mass, SFR, sSFR, and high-frequency radio spectral index ($\alpha_{\rm 1.3GHz}^{\rm 3\,GHz}$) bins. Our stacking analyses show that the averaged radio spectral index also steepens with increasing stellar mass and flattens with sSFR of SFGs. In addition, the radio spectral indices at low and high frequencies have a positive correlation.  

5. We use the two different spectral indices measured at low (0.33--1.3\,GHz) and high frequencies (1.3--3\,GHz) to $k$-correct the observed flux density at MeerKAT 1.3\,GHz, and thus obtain two sets of $q_{\rm IR}$ values for the SFGs in our sample. For the SFGs with measured flux density at the three radio frequencies, using the high-frequency spectral index for $k$-correction underestimates the $q_{\rm IR}$ for  >17\% of them that have a flatter radio spectrum at low frequency. For all of the 3\,GHz-selected SFGs, using the spectral index at high frequency in $k$-correction slightly overestimate the evolution of $q_{\rm IR}$ with cosmic time, although both choices of the radio spectral index (at low or high frequency) result in similar trends for $q_{\rm IR}$-redshift and $q_{\rm IR}$-stellar mass relations. Therefore, deep low-frequency radio continuum data are essential for an accurate study of FIRRC of SFGs.

\section*{Acknowledgements}
We are grateful to the anonymous referee for a detailed report and valuable comments that improved the quality of this work.
FXA acknowledges financial support from the Inter-University Institute for Data Intensive Astronomy (IDIA). IDIA is a partnership of the University of Cape Town, the University of Pretoria and the University of the Western Cape. IRS acknowledges support from STFC (ST/T000244/1). IHW acknowledges support from the Oxford Hintze Centre for Astrophysical Surveys which is funded through generous support from the Hintze Family Charitable Foundation. CLH acknowledges support from the Leverhulme Trust through an early career research fellowship. SJ acknowledges financial support from the Spanish Ministry of Science, Innovation and Universities (MICIU) under grant AYA2017-84061-P, co-financed by FEDER (European Regional Development Funds) and the Agencia Estatal de Investigaci\'{o}n del Ministerio de Ciencia e Innovaci\'{o}n (AEI-MCINN) under grant (La evoluci\'{o}n de los c\'{u}mulos de galaxias desde el amanecer hasta el mediod\'{i}a c\'{o}smico) with reference (PID2019-105776GB-I00/DOI:10.13039/501100011033). YA acknowledges financial support by NSFC grant 11933011 and the science research grant from the China Manned Space Project with NO. CMS-CSST-2021-B06. SMR hereby acknowledged the financial assistance of the National Research Foundation (NRF) towards this research. Opinions expressed and conclusions arrived at, are those of the author and are not necessarily to be attributed to the NRF.
 
We acknowledge the use of the ilifu cloud computing facility - www.ilifu.ac.za, a partnership between the University of Cape Town, the University of the Western Cape, the University of Stellenbosch, Sol Plaatje University, the Cape Peninsula University of Technology and the South African Radio Astronomy Observatory.  The ilifu facility is supported by contributions from the IDIA, the Computational Biology division at UCT and the Data Intensive Research Initiative of South Africa (DIRISA).
 
The MeerKAT telescope is operated by the South African Radio Astronomy Observatory, which is a facility of the National Research Foundation, an agency of the Department of Science and Innovation. We acknowledge use of the IDIA data intensive research cloud for data processing.  The authors acknowledge the Centre for High Performance Computing (CHPC), South Africa, for providing computational resources to this research project.

%%%%%%%%%%%%%%%%%%%%%%%%%%%%%%%%%%%%%%%%%%%%%%%%%%
\section*{Data Availability}

The MeerKAT 1.3\,GHz data were accessed from the South African Radio Astronomy Observatory (SARAO, www.ska.ac.za). The derived data generated in this research will be shared on reasonable request to the corresponding author.



\bibliographystyle{mnras}
\bibliography{MIGHTEE_radio_spectra_FRC_Fangxia} % if your bibtex file is called example.bib

%%%%%%%%%%%%%%%%%%%%%%%%%%%%%%%%%%%%%%%%%%%%%%%%%%

% Don't change these lines
\bsp	% typesetting comment
\label{lastpage}
\end{document}